\documentclass[aps,prc,reprint,superscriptaddress, nofootinbib]{revtex4-2}

\usepackage[T1]{fontenc}
\usepackage[utf8]{inputenc}   % Overleaf default
\usepackage{lmodern}
\usepackage{microtype}

\usepackage{amsmath,amssymb,amsthm,mathtools,bm}
\usepackage{siunitx}
\usepackage{booktabs}
\usepackage{graphicx}
\usepackage{placeins}
\usepackage{xcolor}

\usepackage{hyperref}
\hypersetup{
  colorlinks=true,
  allcolors=blue
}
\usepackage[capitalize,noabbrev]{cleveref}

\newcommand{\Normal}{\mathcal{N}}
\newcommand{\Gam}{\mathrm{Gamma}}
\newcommand{\Pois}{\mathrm{Poisson}}

\newcommand{\tr}{\mathrm{T}}

\begin{document}

\title{\boldmath Empirical-Bayes Unfolding of \texorpdfstring{$\gamma$}{gamma}-ray Spectra}

\author{A.~H.~Mj{\o}s}
\affiliation{Department of Physics, University of Oslo, N-0316 Oslo, Norway}
\affiliation{Norwegian Nuclear Research Centre, Norway}

\author{E.~Lima}
\affiliation{Department of Physics, University of Oslo, N-0316 Oslo, Norway}
\affiliation{Norwegian Nuclear Research Centre, Norway}

\author{A.~Kvellestad}
\affiliation{Department of Physics, University of Oslo, N-0316 Oslo, Norway}

\author{A.~C.~Larsen}
\affiliation{Department of Physics, University of Oslo, N-0316 Oslo, Norway}
\affiliation{Norwegian Nuclear Research Centre, Norway}

\author{M.~Hjorth-Jensen}
\affiliation{Department of Physics, University of Oslo, N-0316 Oslo, Norway}
\affiliation{Center for Computing in Science Education, University of Oslo, N-0316 Oslo, Norway}

\date{\today}  

\begin{abstract}
Unfolding observed $\gamma$-ray spectra is an ill-conditioned Poisson inverse problem. Detector response effects and finite energy resolution make distinct non-negative emitted $\gamma$-ray spectra nearly indistinguishable after forward mapping, so direct inversion can strongly amplify statistical fluctuations. Here, we present an empirical-Bayes hierarchical unfolding method that preserves the Poisson counting structure, enforces non-negativity, and incorporates background through a joint ON/OFF likelihood. The prior on the emitted spectrum is centered on an automatically selected Richardson-Lucy reference spectrum, with an adaptive width that remains broad in weakly constrained regions. Posterior inference is performed with the No-U-Turn Sampler, and simultaneous uncertainty bands are reported for the resolution-limited unfolded spectrum. Our Bayesian method provides a robust and extensible framework for uncertainty quantification in unfolding, and a direct comparison with a recent frequentist regularized maximum-likelihood method gives highly consistent unfolded spectra in representative high- and low-statistics cases.
\end{abstract}

\maketitle
\raggedbottom

\section{Introduction}
\label{sec:intro}

The extraction of nuclear statistical properties, such as the nuclear level density (NLD) and the $\gamma$-ray strength function ($\gamma$SF), is central to nuclear-structure studies and astrophysical $(n,\gamma)$ reaction-rate calculations~\cite{larsen201969, schiller2000extraction,toft2010sn}. Within the Oslo method, these properties are derived from primary $\gamma$-ray spectra extracted through a multi-step analysis of particle-$\gamma$-ray coincidence data. The excitation energy $E_x$ is determined using a particle-telescope array such as SiRi~\cite{guttormsen2011168}, while the coincident $\gamma$-rays are measured with a high-efficiency spectrometer such as the OSCAR scintillator array~\cite{zeiser2020oscar}. The coincidence data are organized as a two-dimensional matrix of observed counts binned in excitation energy $E_x$ and $\gamma$-ray energy $E_\gamma$. For a fixed excitation-energy bin, the experimental observable is a $\gamma$-ray spectrum of registered counts in the $E_\gamma$ bins. 

The task of the unfolding is to infer the corresponding emitted spectrum while accounting for counting statistics, detector response effects, and background. This task is naturally formulated as a Poisson inverse problem. However, the discretized forward operator is typically ill-conditioned because the detector response and finite energy resolution redistribute counts and limit the recoverable spectral detail. As a result, different non-negative emitted spectra can produce nearly indistinguishable detected spectra, and direct inversion may amplify Poisson fluctuations along weakly constrained directions in the solution space of possible emitted spectra. These features motivate unfolding methods that combine regularization with uncertainty quantification while preserving the non-negative counting structure of the data.

Unfolding and detector-response correction have a long history in nuclear spectroscopy beyond the Oslo method. In high-spin $\gamma$-ray spectroscopy, response correction has long been used to extract physical information from continuum spectra~\cite{diamond1980highspin}. Dedicated stripping procedures based on measured Ge-detector response functions have been developed for source and in-beam spectra, singles spectra, and $\gamma$--$\gamma$ coincidence matrices~\cite{radford1987stripping,love1989unfolding}. Alternative approaches have also been explored, including neural-network unfolding of photon and neutron spectra measured with liquid scintillators~\cite{koohifayegh1993neural}.

Related unfolding problems have also been studied extensively in high-energy physics, where the goal is likewise to infer an underlying spectrum from counts redistributed by a detector response. D'Agostini's iterative Bayesian unfolding is widely used in that field, and its relation to the Richardson-Lucy family of expectation-maximization algorithms has been discussed explicitly~\cite{dagostini1995unfolding,zech2013rl}. Fully Bayesian and empirical-Bayes formulations have also been developed for unfolding in high-energy physics~\cite{choudalakis2012fbu,kuusela2015unfolding}.

More recently, a frequentist regularized maximum-likelihood estimation (RMLE) framework was developed for unfolding $\gamma$-ray spectra in nuclear-physics applications, with explicit non-negativity constraints, response-aware forward modeling, and bootstrap-based uncertainty quantification~\cite{lima2025}. This framework provides the most direct benchmark for the Bayesian approach developed here. In our formulation, regularization is expressed through a prior, uncertainty is represented by the posterior distribution, and model adequacy can be examined through prior and posterior predictive simulations. We assess the method on controlled synthetic spectra, study its sensitivity to the empirical-Bayes prior construction, and compare the resulting posterior summaries with the RMLE approach.

The Bayesian model combines the factorized Oslo-method response with a Poisson likelihood, a non-negative prior on the emitted spectrum $\bm{\mu}$, and a joint ON/OFF model for signal-plus-background and background-only counts. The joint counting model avoids deterministic background subtraction and preserves the non-negative Poisson structure in low-count regimes. We do not use the posterior over $\bm{\mu}$ itself as the primary reported result. The emitted spectrum can contain high-frequency components that are only weakly constrained by the detector response, and posterior draws in $\bm{\mu}$ may therefore contain substantial sub-resolution variation. Instead, we report inference for the resolution-limited spectrum $\bm{\eta} = \mathbf{G}_\gamma \bm{\mu}$, where $\mathbf{G}_\gamma$ is the $\gamma$-ray energy-resolution operator. This maps each posterior draw to the detector-resolution scale before the uncertainty band is constructed. The use of $\bm{\eta}$ does not remove the ill-posedness of the unfolding problem, but it places the posterior draws, truth spectra, and uncertainty bands on the detector-resolution scale used for reporting.

The main modeling challenge is the construction of a prior that stabilizes the inverse problem without biasing physically relevant solutions. Highly flexible hierarchical priors are appealing in principle, but in this inverse problem they can produce posterior geometries that are difficult to sample reliably, especially when the emitted spectrum contains narrow peak-like structures. In contrast, strong smoothness priors can make posterior sampling stable while biasing the unfolded spectrum. We therefore use an empirical-Bayes construction. A background-aware Richardson-Lucy (RL) estimate is used only to define a data-driven reference scale and an adaptive prior width. Conditional on this empirical prior construction, the emitted spectrum and background are inferred with a Bayesian hierarchical model.

Posterior sampling is performed using the No-U-Turn Sampler (NUTS)~\cite{hoffman2014nuts}, implemented in \texttt{PyMC}~\cite{pymc2023}. The unfolding is carried out independently for each excitation-energy bin, so the computational task is a sequence of one-dimensional Poisson inverse problems. For the four-chain configuration used here, with 2000 warm-up iterations and 2000 posterior draws per chain, the wall-clock times for the benchmarked spectra range from roughly one minute to about half an hour per spectrum on a multi-core laptop CPU, depending on the number of $E_\gamma$ bins, the counting statistics, and the sampler backend.

The remainder of this article is organized as follows. Section~\ref{sec:theory} defines the factorized forward response and summarizes the mechanisms that make the unfolding ill-conditioned. Section~\ref{sec:method} presents the empirical-Bayes hierarchical prior, the ON/OFF background model, the NUTS-based posterior sampling scheme, and the simultaneous uncertainty summaries used in the analysis. Section~\ref{sec:results} validates the method on synthetic spectra with known truth, including prior-to-posterior contraction, posterior predictive checks, sensitivity to the empirical-Bayes construction, robustness to alternative modeling choices, low-statistics prior dependence, and a direct comparison with the frequentist RMLE framework. Conclusions are given in Sec.~\ref{sec:conclusion}, followed by an outlook in Sec.~\ref{sec:outlook}.

\section{Theory}
\label{sec:theory}

\subsection{Forward model and counting statistics}
\label{subsec:forward}

\subsubsection{Notation and conventions}
We represent the emitted spectral intensity as a matrix $\mathbf{M} \in \mathbb{R}_+^{J \times K}$, where the indices $j \in \{1,\dots,J\}$ and $k \in \{1,\dots,K\}$ correspond to the discretized $\gamma$-energy $E_\gamma$ and excitation energy $E_x$, respectively. Under this convention, each column vector $\bm{\mu}_k \in \mathbb{R}_+^J$ represents the emitted $\gamma$-ray spectrum for a fixed excitation-energy bin $k$. The corresponding expectation of detected signal counts is denoted by the matrix $\mathbf{V} \in \mathbb{R}_+^{J \times K}$. Operators modeling the detector response along the $\gamma$-energy axis act on the $E_\gamma$ index and therefore multiply $\mathbf{M}$ from the left. Operators modeling resolution effects along the excitation-energy axis act on the $E_x$ index and therefore multiply from the right. The notation is summarized in Table~\ref{tab:notation}.

\paragraph*{Matrix-display convention.}
In matrix figures, we place $E_\gamma$ on the horizontal axis and $E_x$ on the vertical axis. This display convention is the transpose of the mathematical notation in Table~\ref{tab:notation}, where global matrices are written as $(E_\gamma, E_x)$ arrays and fixed-$E_x$ spectra are columns.

\begin{table}[htbp]
\caption{Summary of the principal quantities. Global quantities are written as $(J \times K)$ matrices on the $(E_\gamma, E_x)$ grid. Local quantities refer to $J$-dimensional spectral vectors for a single excitation-energy bin $k \in \{1,\dots,K\}$.}
\label{tab:notation}
\begin{ruledtabular}
\begin{tabular}{llll}
Quantity & Dim. & Global & Local \\
\hline
Emitted intensity             & $J \times K$ & $\mathbf{M}$ & $\bm{\mu}$ \\
Resolution-limited intensity        & $J \times K$ & $\mathbf{H}$ & $\bm{\eta}$ \\
Expected detected signal               & $J \times K$ & $\mathbf{V}$ & $\bm{\nu}$ \\
Expected background           & $J \times K$ & $\mathbf{B}$ & $\mathbf{b}$ \\
Observed ON counts              & $J \times K$ & $\mathbf{N}$ & $\mathbf{n}$ \\
Observed OFF counts            & $J \times K$ & $\mathbf{N}_{\mathrm{off}}$ & $\mathbf{n}_{\mathrm{off}}$ \\
\hline
$\gamma$-energy response        & $J \times J$ & $\mathbf{R}_\gamma$ & --- \\
$\gamma$-energy redistribution       & $J \times J$ & $\mathbf{D}$ & --- \\
$\gamma$-energy resolution    & $J \times J$ & $\mathbf{G}_\gamma$ & --- \\
Excitation-energy resolution         & $K \times K$ & $\mathbf{G}_{\mathrm{in}}$ & --- \\
\end{tabular}
\end{ruledtabular}
\end{table}

\subsubsection{Response factorization}
The forward mapping $\mathbf{M} \mapsto \mathbf{V}$ is expressed as a composition of three linear operators applied in a physically motivated sequence. These operators represent the specific instrumental responses of the SiRi particle detectors and the OSCAR scintillators used in the Oslo method:
\begin{itemize}
  \item $\mathbf{G}_{\mathrm{in}} \in \mathbb{R}_+^{K \times K}$ is a symmetric broadening matrix modeling the finite excitation-energy resolution of the SiRi detectors. It multiplies $\mathbf{M}$ from the right.
  \item $\mathbf{D} \in \mathbb{R}_+^{J \times J}$ is a lower-triangular redistribution matrix encoding the physical interaction effects, such as Compton scattering and pair-production escape peaks, within the OSCAR crystals. It acts from the left on $\mathbf{M}$.
  \item $\mathbf{G}_\gamma \in \mathbb{R}_+^{J \times J}$ is a resolution-broadening matrix modeling the instrumental $\gamma$-energy resolution of the OSCAR array. It acts on $\mathbf{M}$ from the left.
\end{itemize}
Crucially, the operators $\mathbf{G}_\gamma$ and $\mathbf{D}$ do not commute ($\mathbf{G}_\gamma \mathbf{D} \neq \mathbf{D} \mathbf{G}_\gamma$), as demonstrated in Ref.~\cite{lima2025}. The factorization order reflects the physical detection sequence in OSCAR: incident photons first undergo energy redistribution $\mathbf{D}$, and the resulting deposited energy is subsequently subject to resolution broadening $\mathbf{G}_\gamma$. With this ordering, we define the resolution-limited intensity matrix $\mathbf{H}$ and the matrix of expected detected-signal counts $\mathbf{V}$ as
\begin{equation}
  \mathbf{H} = \mathbf{G}_\gamma \mathbf{M} \mathbf{G}_{\mathrm{in}},
  \qquad
  \mathbf{V} = \mathbf{G}_\gamma \mathbf{D} \mathbf{M} \mathbf{G}_{\mathrm{in}}.
  \label{eq:forward}
\end{equation}
In this notation, the composite $\gamma$-energy response operator for OSCAR is
\begin{equation}
  \mathbf{R}_\gamma := \mathbf{G}_\gamma \mathbf{D} \in \mathbb{R}_+^{J \times J}.
  \label{eq:Rgamma}
\end{equation}
For a fixed excitation-energy bin and with $\mathbf{G}_{\mathrm{in}}=\mathbf{I}$, the local forward quantities are 
\begin{equation}
\bm{\eta} = \mathbf{G}_\gamma \bm{\mu}, \qquad \bm{\nu}= \mathbf{R}_\gamma \bm{\mu} =  \mathbf{G}_\gamma \mathbf{D} \bm{\mu}.
\label{eq:local-forward}
\end{equation}
The operators are normalized to preserve the total expected counts along the corresponding axes. Setting $\mathbf{G}_{\mathrm{in}} = \mathbf{I}$, the normalization of $\mathbf{R}_\gamma$ ensures
\begin{equation}
  \sum_{j=1}^J V_{j,k} = \sum_{j=1}^J M_{j,k},
  \label{eq:count-preserve}
\end{equation}
where $V_{j,k}$ and $M_{j,k}$ denote the elements of the respective matrices.

\subsubsection{Poisson statistics}
Conditioned on the expected detected signal $\mathbf{V}$ and the expected background $\mathbf{B}$, the observed ON counts $\mathbf{N}$ are modeled as independent Poisson variables across the active $(E_\gamma, E_x)$ grid,
\begin{equation}
  N_{j,k} \mid V_{j,k}, B_{j,k} \sim \mathrm{Poisson}(V_{j,k} + B_{j,k}).
  \label{eq:poisson}
\end{equation}
When a separate background measurement is available, the corresponding OFF counts are modeled as 
\begin{equation}
  N_{\mathrm{off},j,k} \mid B_{j,k}
  \sim
  \mathrm{Poisson}(B_{j,k}).
  \label{eq:off-poisson}
\end{equation}
This joint ON/OFF likelihood preserves the non-negative Poisson counting structure and avoids deterministic background subtraction. The background model used for inference is described in Sec.~\ref{subsec:bgmodel}.

\subsubsection{Physical content of \texorpdfstring{$\mathbf{D}$}{D}}
For the inorganic LaBr$_3$(Ce) scintillators used in the OSCAR array~\cite{zeiser2020oscar}, the matrix $\mathbf{D}$ summarizes the energy-redistribution part of the detector response. It describes how physical energy-deposition processes redistribute the incident photon energy across the detected $E_\gamma$ bins. This includes the full-energy peak events, the Compton continuum arising from partial energy deposition, and escape features (single and double escape peaks) that appear once the pair-production threshold is exceeded. A detailed discussion of these detector-physics mechanisms is provided in standard references such as Ref.~\cite{knoll2010rdm}.

\subsection{Why inversion is hard: ill-conditioning and noise amplification}
\label{subsec:illcond}

Unfolding aims to recover a non-negative emitted spectrum from noisy counts after the action of a detector response operator. The fundamental difficulty is structural. Detector redistribution and resolution broadening attenuate some spectral directions much more strongly than others, so different emitted spectra can produce nearly indistinguishable detected spectra. After discretization, this appears as a response matrix with rapidly decaying singular values and weakly informed directions in the solution space~\cite{engl1996regularization,hansen1998rankdeficient,hansen2010rank}.

For a fixed excitation-energy bin $k \in \{1,\dots,K\}$, we set $\mathbf{G}_{\mathrm{in}}=\mathbf{I}$ to focus on the local one-dimensional $E_\gamma$ unfolding problem. The detected-signal forward mapping then reduces to
\begin{equation}
  \bm{\nu} = \mathbf{R}_\gamma \bm{\mu},
  \label{eq:local-detected-forward}
\end{equation}
where $\bm{\mu}$ and $\bm{\nu}$ are the $J$-dimensional emitted and detected-signal expectation vectors. Here $\mathbf{R}_\gamma = \mathbf{G}_\gamma \mathbf{D}$ is the composite response from Eq.~\eqref{eq:Rgamma}.

\subsubsection{Singular values and numerical rank}
Using the standard singular value decomposition (SVD) notation~\cite{golub2013matrix}, the response matrix can be written as
\begin{equation}
  \mathbf{R}_\gamma = \mathbf{U}\bm{\Sigma}\mathbf{W}^{\tr},
  \qquad \bm{\Sigma}= \mathrm{diag}(\sigma_1,\dots,\sigma_J),
\end{equation}
with singular values
\begin{equation}
  \sigma_1\ge \sigma_2\ge \cdots \ge \sigma_J \ge 0.
\end{equation}
The existence of singular values should not be confused with singularity of the matrix. The response matrix is exactly rank deficient only if one or more of these singular values are zero. In that case, the exact rank $r$ is the number of non-zero singular values, and the nullspace is spanned by the right singular vectors corresponding to $\sigma_\ell=0$.

In practice, a response matrix may be technically full rank while still being numerically close to rank deficient. Smoothing and redistribution operators typically lead to singular values that decay rapidly after discretization, so directions associated with small $\sigma_\ell$ have only a weak effect on the detected spectrum~\cite{engl1996regularization,hansen1998rankdeficient,hansen2010rank}. For a full-rank square matrix, the two-norm condition number is 
\begin{equation}
  \kappa_2(\mathbf{R}_\gamma)=\frac{\sigma_1}{\sigma_J}.
\end{equation}
If $\sigma_J=0$, this condition number is infinite. When the smallest singular values are non-zero but below the numerical or statistical resolution of the problem, it is more useful to speak of an effective numerical rank $r$ and an effective condition number $\sigma_1/\sigma_r$ for the retained singular directions.

\subsubsection{Noise amplification and the discrete Picard condition}
The effect of small singular values is most transparent in the linear least-squares analogue of the unfolding problem. If the background is fixed and $\bm{\nu}$ denotes the expected detected-signal vector, the least-squares solution is 
\begin{equation}
  \bm{\mu}_{\mathrm{LS}}=\mathbf{R}_\gamma^\dagger \bm{\nu},
\end{equation}
where $\mathbf{R}_\gamma^\dagger$ is the Moore-Penrose pseudoinverse. A perturbation $\Delta\bm{\nu}$ in the detected spectrum gives
\begin{equation}
  \Delta\bm{\mu}_{\mathrm{LS}}= \mathbf{R}_\gamma^\dagger \Delta\bm{\nu} =
  \sum_{\ell=1}^{r} \frac{\mathbf{u}_\ell^\tr\Delta\bm{\nu}}{\sigma_\ell}
  \mathbf{w}_\ell,
  \label{eq:svd-backproj-illcond}
\end{equation}
where $\mathbf{u}_\ell$ and $\mathbf{w}_\ell$ are the left and right singular vectors, respectively. Hence, a perturbation component along $\mathbf{u}_\ell$ is amplified by $1/\sigma_\ell$ in the corresponding solution direction $\mathbf{w}_\ell$. Small singular values therefore make the inverse highly sensitive to statistical fluctuations.

A standard characterization of discrete ill-posedness is the discrete Picard condition. In the absence of noise, stable inversion requires the singular vector coefficients of the exact detected spectrum to decay sufficiently fast relative to the singular values. Once the coefficients $\mathbf{u}_\ell^\tr\bm{\nu}$ are dominated by statistical noise, division by small $\sigma_\ell$ makes the tail of the SVD expansion unstable~\cite{hansen1998rankdeficient,hansen2010rank}. Classical regularization methods stabilize the inversion by damping or removing these weakly informed singular directions. An overview is provided in Sec.~\ref{subsec:classical-reg}.

\subsubsection{Approximate nullspaces and resolution-limited spectra}
If $\mathbf{R}_\gamma$ is rank deficient, the inverse problem is non-unique even for noise-free data. If $\bm{\mu}_\star$ is one solution, then
\begin{equation}
  \bm{\mu} = \bm{\mu}_\star + \sum_{\ell=r+1}^{J} c_\ell \mathbf{w}_\ell,
  \label{eq:nullspace-illcond}
\end{equation}
is also a solution for arbitrary coefficients $c_\ell$, because the right singular vectors $\mathbf{w}_\ell$ with $\sigma_\ell=0$ span $\mathrm{null}(\mathbf{R}_\gamma)$. Even when the response matrix is full rank, very small singular values create an approximate nullspace. The emitted spectrum can change substantially along the corresponding right singular directions while producing only a small change in the expected detected counts. Under finite counting statistics, such changes may be statistically indistinguishable from Poisson fluctuations~\cite{engl1996regularization,hansen1998rankdeficient,hansen2010rank}.

This practical non-uniqueness is tied to the finite energy resolution and redistribution of the detector response. High-frequency features in the emitted spectrum are strongly attenuated by the resolution operator $\mathbf{G}_\gamma$, so they can be weakly constrained by the data even when they have large amplitude in $\bm{\mu}$-space. For this reason, the posterior summaries in this work are reported for the resolution-limited spectrum $\bm{\eta}$ rather than for the unresolved emitted spectrum alone. This does not remove the ill-posedness of the unfolding problem, but it places the reported uncertainty on the detector-resolution scale used in the forward model.

\subsubsection{Signal-dependent noise and background subtraction}
Unlike additive Gaussian noise, Poisson noise is signal-dependent. With background included, 
\begin{equation}
  \mathrm{Var}(n_j \mid \nu_j,b_j) = \nu_j+b_j.
\end{equation}
In low-count bins, the relative fluctuations are large, further weakening the identifiability of spectral directions already attenuated by the response $\mathbf{R}_\gamma$. Conditioned on a fixed background expectation, the local curvature of the ON-count log-likelihood is quantified by the expected Fisher information matrix
\begin{equation}
  \bm{\mathcal{I}}(\bm{\mu}) = \mathbf{R}_\gamma^\tr \mathrm{diag}(\mathbf{R}_\gamma \bm{\mu} + \mathbf{b})^{-1} \mathbf{R}_\gamma,
  \label{eq:fisher-x-illcond}
\end{equation}
where $\mathbf{b}$ is the background expectation for the selected excitation-energy bin. In the absence of background, $\mathbf{b}=0$. The information matrix reflects the ill-conditioning of $\mathbf{R}_\gamma$. Directions that produce only small changes in the expected detected signal are therefore weakly constrained by the likelihood.

When a separate background measurement is available, naive subtraction replaces the observed total counts $\mathbf{n}$ with $\mathbf{n}-\mathbf{n}_{\mathrm{off}}$. This difference is not Poisson-distributed. For equal ON/OFF exposure factors it follows a Skellam distribution and can yield negative values~\cite{skellam1946frequency}. In the same equal-exposure case, with 
\begin{equation}
    n_j \sim \mathrm{Poisson}(\nu_j+b_j),
    \qquad n_{\mathrm{off},j}\sim \mathrm{Poisson}(b_j),
\end{equation}
the subtraction estimator is unbiased,
\begin{equation}
    \mathbb{E}[n_j-n_{\mathrm{off},j}]=\nu_j,
\end{equation}
but has variance 
\begin{equation}
  \mathrm{Var}(n_j-n_{\mathrm{off},j}) = \nu_j+2b_j.
\end{equation}
The two background terms arise because background fluctuations are present both in the ON measurement and in the independent OFF measurement. Using the standard deviation of the subtraction estimator as an uncertainty scale, its relative uncertainty is $\sqrt{\nu_j+2b_j}/{\nu_j}$. This quantity becomes large when the signal is small compared with the background, and it diverges as $\nu_j\to0$. Consequently, fluctuations with $n_j<n_{\mathrm{off},j}$ can occur even though the signal expectation is non-negative. This motivates the joint ON/OFF likelihood formulation used in Sec.~\ref{sec:method}.

\subsection{Classical regularization and its limitations}
\label{subsec:classical-reg}

Classical regularization stabilizes the noise-amplification mechanisms described in Sec.~\ref{subsec:illcond} by trading variance for bias through explicit penalties or early stopping. A common frequentist approach adds a penalty term to the Poisson negative log-likelihood and solves the resulting optimization problem subject to the physical constraint $\bm{\mu} \ge 0$. Defining the negative log-likelihood for a single excitation-energy bin, up to constants independent of $\bm{\mu}$, as
\begin{equation}
\begin{split}
  \Phi(\bm{\mu},\mathbf b) &= -\log \mathcal{L}(\bm{\mu},\mathbf b) \\
                           &= \sum_{j=1}^{J} \left[ \nu_j + b_j - n_j\log(\nu_j+b_j) \right],
\end{split}
\label{eq:poisson-nll}
\end{equation}
with $b_j=0$ when background is absent, a Tikhonov-type estimator solves 
\begin{equation}
  \widehat{\bm{\mu}}_{\lambda} = \arg\min_{\bm{\mu} \ge 0}
  \left\{ \Phi(\bm{\mu},\mathbf b) + \lambda \|\mathbf{L}\bm{\mu}\|_2^2 \right\},
  \label{eq:tikhonov}
\end{equation}
where $\mathbf{L}$ determines the nature of the regularization. An identity matrix $\mathbf{L}=\mathbf{I}$ yields ridge-type shrinkage, while first- or second-order finite-difference operators favor spectral smoothness. The hyperparameter $\lambda > 0$ governs the trade-off between the data fit and the stability of the solution~\cite{tikhonov1977solutions,hansen2010rank}.

For Poisson models, iterative expectation-maximization schemes, such as the Richardson-Lucy algorithm~\cite{richardson1972rl,lucy1974rl} and the Shepp-Vardi algorithm~\cite{shepp1982em}, maximize the likelihood under the constraint $\bm{\mu} \ge 0$. In practical Oslo-method applications, these methods are often regularized implicitly by early stopping, and in some variants by additional smoothing constraints~\cite{guttormsen1996371,midtbo2021ompy}. While these techniques can recover sharp features when properly tuned, they face two important challenges. First, selecting an appropriate regularization strength $\lambda$, or an appropriate stopping iteration, can be sensitive and may vary between different datasets. Second, uncertainty quantification around a penalized or early-stopping point estimate is non-trivial. Gaussian error bars around such an estimate can underestimate uncertainty near the non-negativity boundary and along weakly informed directions of the response $\mathbf{R}_\gamma$. These limitations motivate the hierarchical Bayesian framework developed in Sec.~\ref{sec:method}.

\subsection{Bayesian regularization and uncertainty}
\label{subsec:bayes-reg}

The Bayesian approach incorporates physical constraints and prior information through probability distributions on the emitted spectrum, and the background when present. For one excitation-energy bin, the posterior has the schematic form
\begin{equation}
  \pi(\bm{\mu}, \mathbf{b} \mid \mathbf{n}, \mathbf{n}_{\mathrm{off}}) \propto \mathcal{L}(\bm{\mu},\mathbf{b})\, \pi(\bm{\mu}) \, \pi(\mathbf{b}).
  \label{eq:bayes-posterior}
\end{equation}
The prior on $\bm{\mu}$ enforces non-negativity and regularizes the weakly identified directions of the inverse problem. If the background is fixed, the maximum-a-posteriori (MAP) estimate satisfies
\begin{equation}
  \widehat{\bm{\mu}}_{\mathrm{MAP}} = \arg\min_{\bm{\mu} \ge 0}
  \left\{ \Phi(\bm{\mu},\mathbf b) - \log \pi(\bm{\mu}) \right\}.
\end{equation}
This shows that the negative log-prior acts as a regularizer analogous to the penalties used in frequentist analysis. 

A key advantage of this framework is that Bayesian inference delivers the posterior distribution rather than only a point estimate. Following the principles of Bayesian inverse-problem theory~\cite{kaipio2005statistical}, this posterior can be used to propagate uncertainty while respecting the physical non-negativity constraint. In ill-conditioned problems, weakly informed directions can appear as broader posterior uncertainty. In the method developed below, the posterior is summarized through the resolution-limited spectrum $\bm{\eta}=\mathbf{G}_\gamma\bm{\mu}$, which places the reported uncertainty on the detector-resolution scale rather than on the unresolved emitted scale. The resulting posterior uncertainty can then be examined through simultaneous uncertainty bands, posterior predictive checks, sensitivity analyses, and direct validation on synthetic data.

\section{Methods}
\label{sec:method}
Although the full detector response is fundamentally a coupled two-dimensional mapping, the Bayesian unfolding developed here is formulated as a one-dimensional problem for a fixed excitation-energy bin. In applications to full Oslo-method matrices, applying the method independently to each excitation-energy bin amounts to neglecting the excitation-energy smearing operator $\mathbf{G}_{\mathrm{in}}$. This is a separate approximation from the one-dimensional unfolding problem itself. Its impact depends on the chosen $E_x$ binning and on how rapidly the spectra vary along the excitation-energy axis. We discuss this approximation in Sec.~\ref{subsec:ex-smear}. In the synthetic validation below, each selected excitation-energy bin is therefore treated as an independent one-dimensional problem. We omit the bin index when referring to the $J$-dimensional spectral vectors in Table~\ref{tab:notation}.

\subsection{Motivation for the empirical-Bayes method}
Before specifying the model in detail, we emphasize that the prior construction is empirical Bayes at its outermost level. In principle, one could assign fully Bayesian hyperpriors to the location and scale parameters of the prior on the emitted spectrum and infer them jointly with the spectrum itself. In the present unfolding problem, this strategy proved to be impractical. The inverse problem is strongly ill-conditioned in the emitted space, $\bm{\mu}$, and the spectra may contain very sharp, nearly $\delta$-like structures occupying only one or a few $E_\gamma$ bins. When additional hierarchical layers are introduced for the prior center and prior width, the posterior geometry becomes substantially more complicated, with weakly identified directions, strong parameter dependence, and poor mixing of gradient-based samplers. During method development, we investigated a broad range of centered and non-centered parameterizations together with several alternative samplers, but did not obtain a formulation that was both computationally reliable and sufficiently accurate in the synthetic validation.

A natural response to these sampling difficulties is to impose stronger structural regularization, for example through smoothness penalties or Gaussian process-type priors. In our tests, such constructions indeed made the posterior easier to sample, but at the cost of introducing visible bias in the unfolded solution. This is especially problematic for Oslo-method spectra, where the physically relevant emitted spectrum in $\bm{\mu}$-space can contain narrow, peak-like structures that are not well represented by globally smooth priors. Since the forward model is defined through the non-commuting operators $\mathbf{G}_\gamma$ and $\mathbf{D}$, smoothing imposed directly in $\bm{\mu}$-space also propagates non-trivially to the reported quantity $\bm{\eta}=\mathbf{G}_\gamma\bm{\mu}$.

We therefore adopt an empirical-Bayes construction. Rather than placing a further hyperprior on the prior center itself, we first compute an RL estimate from the observed spectrum~\cite{richardson1972rl,lucy1974rl} and use it only as a data-driven reference for the local scale of the emitted intensity. This reference is not interpreted as a truth estimate. In particular, we use a semi-converged RL reference, where the iteration is selected by comparing the deterministic change in the resolution-limited RL spectrum with a Poisson-resampling noise level. This gives a reference that captures the approximate location of high- and low-intensity regions without treating late-iteration RL fluctuations as physical structures. The role of the RL estimate is therefore limited to defining a stability-floored prior center and an adaptive prior width that distinguishes bins where the data are expected to be highly informative from those where the likelihood is weak and greater prior flexibility is required.

Conditioned on this data-driven prior construction, inference for the emitted spectrum and the background remains fully Bayesian. What is not propagated is the uncertainty associated with the prior-setting step itself. We regard this as an explicit and necessary approximation: in synthetic validation, it yields posterior distributions that are substantially more reliable than those from the more ambitious fully Bayesian alternatives considered during development. The sensitivity analyses in Sec.~\ref{subsec:results-sensitivity} quantify how strongly the resulting posterior depends on this empirical-Bayes choice.

\subsection{Generation of synthetic data}
\label{subsec:synthetic}

\begin{figure*}
  \centering
  \includegraphics[width=\textwidth]{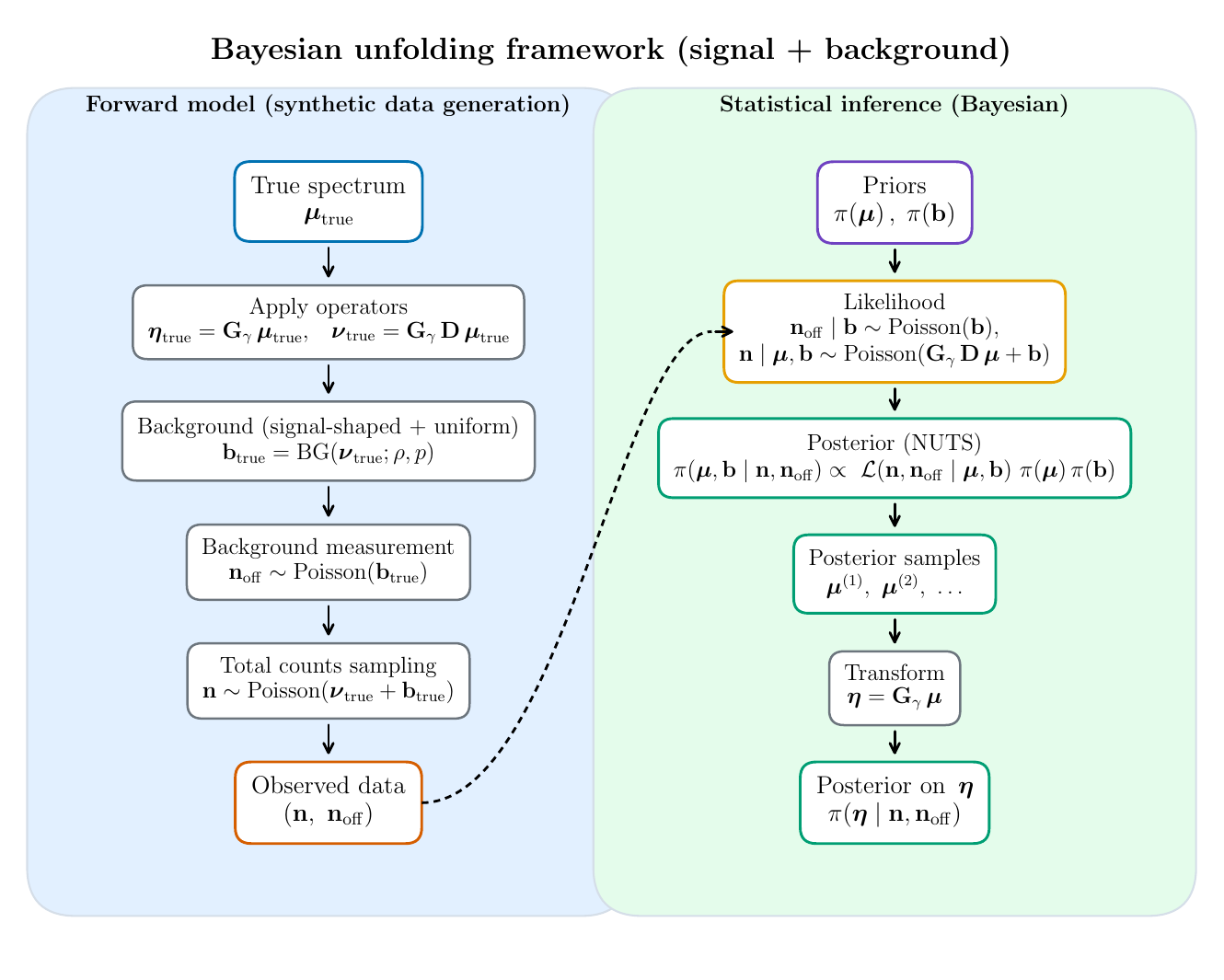}%
  \caption{End-to-end workflow for synthetic-data generation and Bayesian unfolding. Left: forward model from emitted truth spectrum $\bm{\mu}_{\mathrm{true}}(E_x,E_\gamma)$ to the resolution-limited spectrum $\bm{\eta}_{\mathrm{true}}=\mathbf{G}_\gamma\bm{\mu}_{\mathrm{true}}$ and the expected detected signal $\bm{\nu}_{\mathrm{true}}=\mathbf{G}_\gamma\mathbf{D}\bm{\mu}_{\mathrm{true}}$, followed by the construction of the background expectation $\mathbf{b}_{\mathrm{true}}$ and Poisson sampling of OFF and ON data.
  Right: Bayesian unfolding model specified in $\bm{\mu}$-space, with priors on $\bm{\mu}$ and $\mathbf{b}$ and a joint ON/OFF Poisson likelihood. Posterior draws are mapped to $\bm{\eta}$ via $\mathbf{G}_\gamma$ for reporting and uncertainty quantification.}
  \label{fig:flowchart}
\end{figure*}

The procedure for synthetic data generation is summarized in the left panel of Fig.~\ref{fig:flowchart}. The emitted truth matrix used in this work was generated with the \texttt{RAINIER} simulation tool~\cite{kirsch201830, zeiser2020rainier} for a representative $^{120}$Sn cascade dataset.\footnote{The weak intensity between the dominant diagonal structures in the all-generation matrix is produced by rare cascades beginning with a very-low-energy $\gamma$-ray transition. The subsequent high-energy transition is emitted from a state slightly below the initial excitation energy and is therefore displaced from the usual diagonal. These events do not appear between the diagonals in the first-generation matrix, which retains only the initial low-energy transition.} This choice of isotope provides realistic Oslo-method-like spectral structure, but the unfolding model and validation procedure are not specific to $^{120}$Sn. In the data-generation step used here, we set $\mathbf{G}_{\mathrm{in}}=\mathbf{I}$. Each selected fixed-$E_x$ spectrum is therefore forward mapped only along the $E_\gamma$ axis through $\mathbf{D}$ and $\mathbf{G}_\gamma$.

The purpose of this synthetic-data study is to develop and validate the one-dimensional unfolding under controlled conditions. The selected fixed-$E_x$ spectra are treated as the emitted truths for the corresponding one-dimensional tests. The $E_x$ rebinning controls the count level and spectral content of these test cases. It is not used to optimize the $\mathbf{G}_{\mathrm{in}}=\mathbf{I}$ approximation for the full $(E_x,E_\gamma)$ matrix. Given $\bm{\mu}_{\mathrm{true}}$, the forward model defined in Eq.~\eqref{eq:forward} yields 
\begin{equation}
  \bm{\eta}_{\mathrm{true}} = \mathbf{G}_\gamma \bm{\mu}_{\mathrm{true}},
  \qquad
  \bm{\nu}_{\mathrm{true}} = \mathbf{G}_\gamma \mathbf{D} \bm{\mu}_{\mathrm{true}}.
\end{equation} 
For each excitation-energy bin, we define the one-dimensional unfolding domain by truncating the $E_\gamma$ axis according to the expected detected-signal spectrum $\bm{\nu}_{\mathrm{true}}$. Specifically, we choose the smallest index $J_{\mathrm{act}}$ such that
\begin{equation}
  \sum_{j=1}^{J_{\mathrm{act}}} \nu_{j,\mathrm{true}}
  \ge (1-\epsilon_{\mathrm{tail}}) \sum_{j=1}^{J_{\mathrm{full}}} \nu_{j,\mathrm{true}},
  \label{eq:eg-domain}
\end{equation}
where $\epsilon_{\mathrm{tail}}$ is a small prescribed tail-mass tolerance. The corresponding energy is denoted $E_\gamma^{\max}$. For brevity, we use $J$ for this active dimension in the following. This choice retains a fixed high fraction of the expected detected-signal mass and automatically incorporates the tails produced by the OSCAR response without imposing a hard cutoff such as $E_\gamma \le E_x$. Since the observed ON counts include Poisson fluctuations and background contributions, $\mathbf{n}$ may remain non-zero beyond this boundary. The boundary is therefore an analysis-domain choice, not a threshold applied to the observed counts.

To model the background, we construct the expected background intensity $\mathbf{b}_{\mathrm{true}}$ from the expected detected-signal spectrum $\bm{\nu}_{\mathrm{true}}$ inside the active unfolding domain. We first normalize the detected-signal shape,
\begin{equation}
  s_j = \frac{\nu_{j,\mathrm{true}}}{\sum_{i=1}^{J}\nu_{i,\mathrm{true}}},
  \qquad j=1,\dots,J.
  \label{eq:bg-shape}
\end{equation}
The total expected background intensity $B$ is set proportional to the total expected detected signal,
\begin{equation}
  B = \rho \sum_{j=1}^{J}\nu_{j,\mathrm{true}},
  \label{eq:bg-total}
\end{equation}
where $\rho \ge 0$ controls the overall background-to-signal ratio. We then split this background into a signal-shaped component and a uniform component. For a uniform fraction $p \in [0,1]$, we define
\begin{equation}
  b_{j,\mathrm{true}}=(1-p)B s_j + p\frac{B}{J}.
  \label{eq:bg-lambda}
\end{equation}
Thus, $p=0$ gives a purely signal-shaped background, while $p>0$ adds a strictly positive uniform component across the analysis bins. With $\bm{\nu}_{\mathrm{true}}$ and $\mathbf{b}_{\mathrm{true}}$ fixed, the observed background and total counts are generated as independent Poisson draws from their respective expectations
\begin{equation}
  \mathbf{n}_{\mathrm{off}} \sim \Pois(\mathbf{b}_{\mathrm{true}}), \quad
  \mathbf{n} \sim \Pois(\bm{\nu}_{\mathrm{true}} + \mathbf{b}_{\mathrm{true}}).
\end{equation}
Because $\bm{\mu}_{\mathrm{true}}$ and $\bm{\eta}_{\mathrm{true}}$ are known, the synthetic data provide a controlled setting for validating the Bayesian unfolding performance.

\begin{figure*}
  \centering  
  \includegraphics{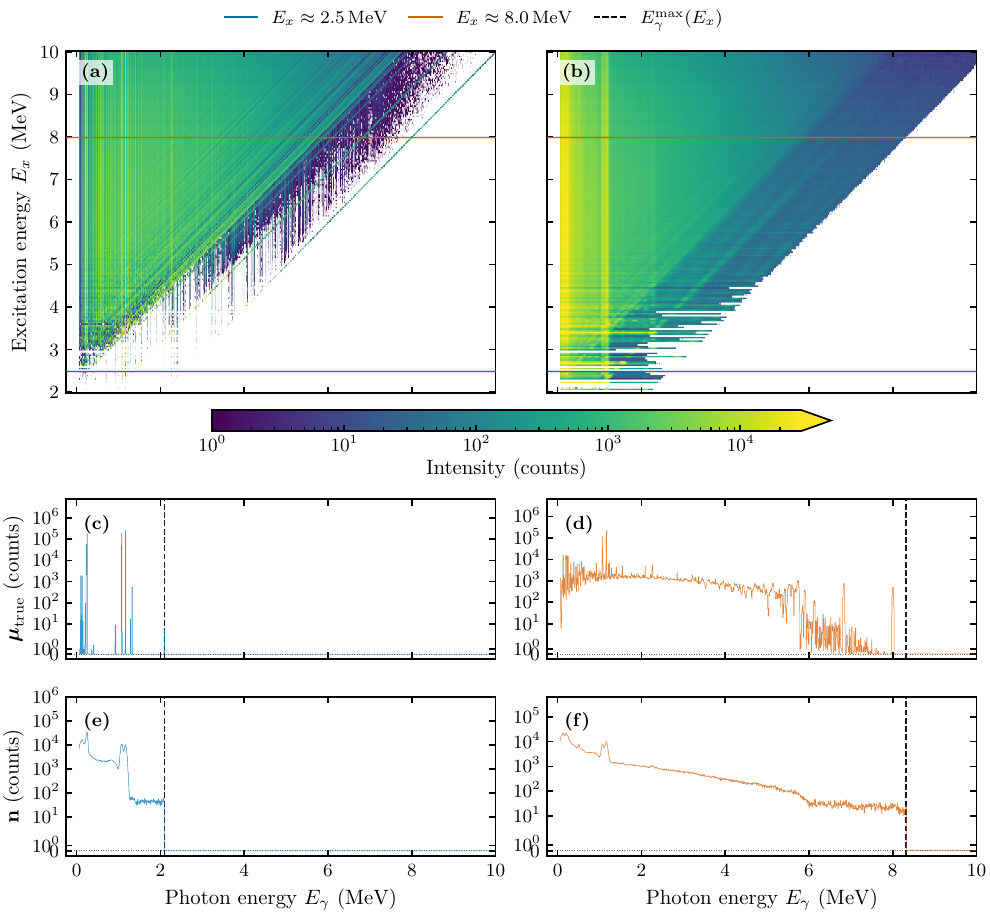}
  \caption{Synthetic $^{120}$Sn RAINIER dataset used for validation of the Bayesian unfolding method. (a) Emitted truth matrix $\mathbf{M}_{\mathrm{true}}$. (b) Observed ON-count matrix $\mathbf{N}$ obtained from $\mathbf{M}_{\mathrm{true}}$ through the forward model steps outlined in Fig.~\ref{fig:flowchart}. The horizontal colored bands indicate the selected excitation-energy bins. Panels (c,d) show representative emitted truth spectra $\bm{\mu}_{\mathrm{true}}$ as functions of $E_\gamma$. Panels (e,f) show the corresponding observed ON-count spectra $\mathbf{n}$ for the same selected $E_x$ bins. The vertical dashed lines mark the excitation-bin dependent upper boundary $E_\gamma^{\max}$ of the unfolding domain, defined from the expected detected-signal mass according to Eq.~\eqref{eq:eg-domain}.}
  \label{fig:synthetic-data}
\end{figure*}

Figure~\ref{fig:synthetic-data} visualizes the synthetic dataset used throughout this work. The mathematical quantities in Table~\ref{tab:notation} are written as $J\times K$ matrices on the $(E_\gamma, E_x)$ grid, where fixed-$E_x$ spectra are columns. In Fig.~\ref{fig:synthetic-data}, the same physical quantities are displayed as matrix images with $E_\gamma$ on the horizontal axis and $E_x$ on the vertical axis. Thus, the matrix panels are transposed relative to the mathematical convention in Table~\ref{tab:notation}. In this displayed orientation, each horizontal slice at fixed $E_x$ corresponds to a one-dimensional $E_\gamma$ spectrum.

The numerical matrix handling and response-matrix specialization are based on \texttt{OMpy}, the Python implementation of the Oslo method~\cite{midtbo2021ompy, ompy_zenodo}. For the OSCAR detector response, we use the \texttt{oscar2020} response set in the OCL response-function repository~\cite{zeiser2020oscar,oclresponse2020}.

\subsection{Bayesian unfolding method}
\label{subsec:bayes}
\begin{figure}
  \centering
  \includegraphics{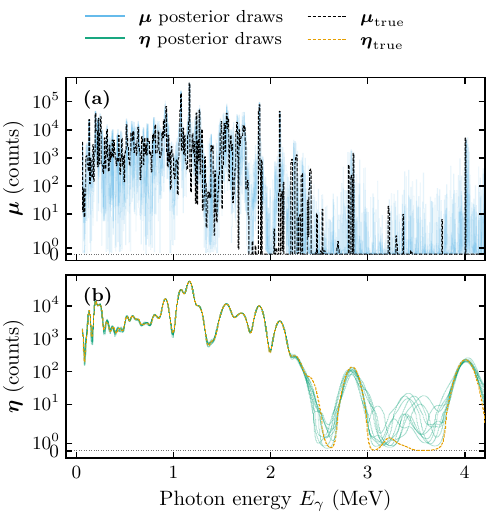}%

  \caption{Ill-posedness of the emitted-space unfolding for a representative excitation-energy bin ($E_x\approx4.0$~MeV). A random subset of 10 posterior draws is shown in both panels. (a) Posterior draws of the emitted spectrum $\bm{\mu}$, together with the emitted truth $\bm{\mu}_{\mathrm{true}}$. (b) The same posterior draws mapped to the resolution-limited spectrum $\bm{\eta}= \mathbf{G}_\gamma\bm{\mu}$, together with $\bm{\eta}_{\mathrm{true}}=\mathbf{G}_\gamma\bm{\mu}_{\mathrm{true}}$. Applying the detector-resolution operator $\mathbf{G}_\gamma$ attenuates much of the high-frequency variation, making $\bm{\eta}$ a more stable quantity to report. Both panels use a symmetric-logarithmic count axis.}
  \label{fig:ill-posedness}
\end{figure}

The right panel of Fig.~\ref{fig:flowchart} summarizes the main components of the Bayesian unfolding model. The $\gamma$-energy resolution is encoded in the smearing kernel $\mathbf{G}_\gamma$, which represents the finite energy resolution of the detector and sets the detector-resolution scale for distinguishable $\gamma$-ray energies. The emitted spectrum $\bm{\mu}$ may contain very sharp, $\delta$-like structures that cannot be fully resolved through unfolding because of this instrumental limit. Figure~\ref{fig:ill-posedness} illustrates the inherent ill-posedness of the problem by comparing posterior draws in the emitted space $\bm{\mu}$ with the same draws mapped to the resolution-limited space $\bm{\eta}$. The emitted-space posterior draws form a broad, highly oscillatory set of curves, reflecting weakly constrained high-frequency directions in $\bm{\mu}$. These variations are largely attenuated after applying the detector-resolution operator $\mathbf{G}_\gamma$. For this reason, although the Bayesian model is specified in $\bm{\mu}$-space, we use $\bm{\eta}$ as the primary quantity for reporting posterior inference.

Priors are specified in $\bm{\mu}$-space for two reasons.  First, as discussed in Sec.~\ref{subsec:forward}, the forward operators $\mathbf{D}$ and $\mathbf{G}_\gamma$ do not commute. A model specified directly in $\bm{\eta}$ would therefore not preserve the physical order of the redistribution and resolution broadening without an additional ill-conditioned inverse mapping back to $\bm{\mu}$. We therefore work with the composite mapping  $\bm{\nu} = \mathbf{G}_\gamma\mathbf{D}\bm{\mu}$. Second, $\bm{\mu}$ represents the intensity of emitted photons per $\gamma$-energy bin without imposed prior correlations. Independent priors across the components of $\bm{\mu}$ are therefore a physically reasonable choice. Posterior correlations across the $E_\gamma$ axis are induced by the structure of the response matrix $\mathbf{R}_\gamma$ and the Poisson likelihood, rather than being imposed by the prior. We also place an independent prior on the latent background expectation $\mathbf{b}$. The joint likelihood for a single excitation-energy bin is defined as the product of two independent Poisson processes,
\begin{equation}
  \begin{split}
  \mathcal{L}(\bm{\mu},\mathbf{b}) = {} & \Pois\bigl(\mathbf{n} \mid \mathbf{G}_\gamma \mathbf{D}\bm{\mu} + \mathbf{b}\bigr) \\
  & \times \Pois\bigl(\mathbf{n}_{\mathrm{off}} \mid \mathbf{b}\bigr).
  \end{split}
  \label{eq:lik}
\end{equation}
Posterior sampling is performed using NUTS (via \texttt{PyMC}~\cite{pymc2023}). The NUTS algorithm is summarized in
Sec.~\ref{subsec:nuts}, while
Sec.~\ref{subsec:sampler-config} specifies the backend and tuning configuration. Convergence diagnostics are described in Sec.~\ref{subsec:diagnostics}.

\subsection{Effect of excitation-energy resolution}
\label{subsec:ex-smear}

For a full experimental Oslo-method matrix, applying the one-dimensional unfolding independently to each excitation-energy bin corresponds to setting $\mathbf{G}_{\mathrm{in}}=\mathbf{I}$, where $\mathbf{G}_{\mathrm{in}}$ describes smearing along the excitation-energy axis. To illustrate the impact of this approximation, we compare the expected detected-signal matrix obtained with and without the excitation-energy smearing operator. In the convention of Sec.~\ref{subsec:forward}, the corresponding bias is 
\begin{equation}
  \Delta\mathbf{V}
  =\mathbf{R}_\gamma \mathbf{M}_{\mathrm{true}}(\mathbf{G}_{\mathrm{in}} - \mathbf{I}).
  \label{eq:ex-smear-bias}
\end{equation}

\begin{figure*}
  \centering
  \includegraphics{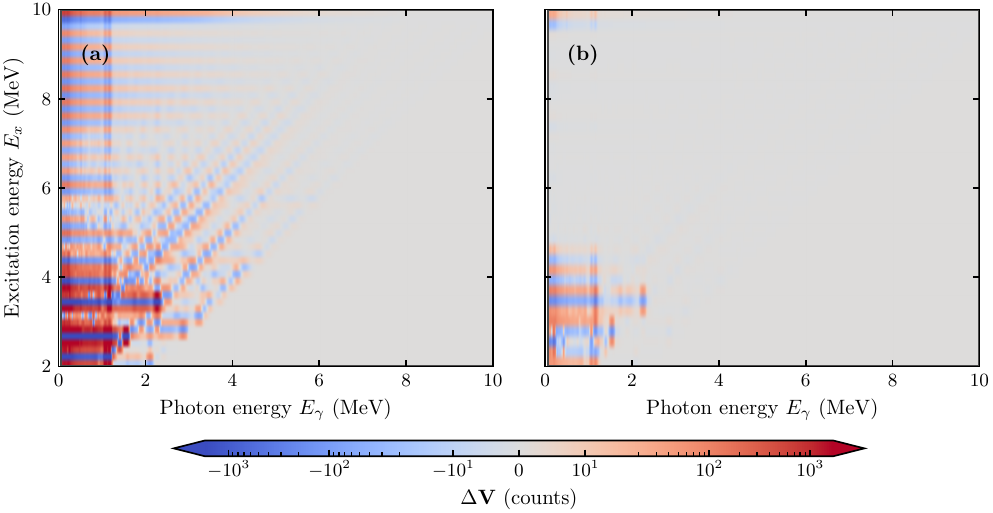}
  \caption{Expected detected-signal bias induced by neglecting excitation-energy smearing in the forward model. Shown is $\Delta\mathbf{V}$ from Eq.~\eqref{eq:ex-smear-bias}, displayed with $E_\gamma$ on the horizontal axis and $E_x$ on the vertical axis. This is the transpose of the mathematical matrix convention used in Table~\ref{tab:notation}. The excitation-energy resolution kernel has $\mathrm{FWHM}=150$~keV in both panels. Panel (a) uses an excitation-energy bin width $\Delta E_x$ of approximately one FWHM, while panel (b) uses $\Delta E_x$ of approximately $1.5$ FWHM. The color scale shows $\Delta\mathbf{V}$ in counts using a symmetric-logarithmic normalization centered at zero.}
  \label{fig:ex-smear}
\end{figure*}

Figure~\ref{fig:ex-smear} illustrates how excitation-energy smearing redistributes counts between neighboring $E_x$ bins, producing an alternating pattern of positive and negative bias along the $E_x$ axis. When neighboring bins are combined into broader $E_x$ bins, opposite-sign contributions partially cancel within the larger bin. This cancellation is visible when going from $\Delta E_x\approx\mathrm{FWHM}$ in panel (a) to $\Delta E_x\approx1.5\,\mathrm{FWHM}$ in panel (b), where the expected detected-signal bias is substantially reduced. The effect is not eliminated entirely, but it becomes much smaller on the scale relevant for the unfolded one-dimensional spectra. For a fixed $\Delta E_x$, the magnitude of $\Delta \mathbf{V}$ is also reduced in the quasi-continuum at higher excitation energies, where the spectral variations along the $E_x$ axis are smoother. The figure therefore illustrates a practical route for applying the fixed-$E_x$ one-dimensional method to full $(E_x, E_\gamma)$ matrices: by using broader $E_x$ bins, the bias from neglecting excitation-energy smearing can be substantially reduced in the spectra passed to the independent one-dimensional unfolding.

\subsection{Richardson-Lucy reference estimate}
\label{subsec:rl}
The empirical-Bayes prior requires a computationally cheap reference spectrum that sets the approximate local scale of the emitted intensity. We use a background-aware RL iteration scheme for this purpose. The Richardson-Lucy algorithm was introduced for Poisson deconvolution problems~\cite{richardson1972rl,lucy1974rl}. In the present work, the RL reference is not used as a final unfolded estimate. It is only used to define the prior center and the adaptive prior-width schedule.

For a fixed excitation-energy bin, let $\mathbf{b}_{\mathrm{ref}}$ denote the fixed background expectation used in the RL iteration. When OFF counts are available, we use the Gamma-Poisson posterior mean derived from the background model in Sec.~\ref{subsec:bgmodel},
\begin{equation}
  b_{\mathrm{ref},j}=\frac{a_0+n_{\mathrm{off},j}}{b_0+1},
\end{equation}
where $a_0$ and $b_0$ are the Gamma prior shape and rate parameters, respectively. If no background is included, $\mathbf{b}_{\mathrm{ref}}=0$. Given the active-domain composite response operator $\mathbf{R}_\gamma=\mathbf{G}_\gamma\mathbf{D}$,\footnote{The \texttt{OMpy} response specialization used here returns the active-domain
response with the standard normalization $\mathbf{R}_\gamma^\tr\mathbf{1}=\mathbf{1}$. Hence, the usual Richardson-Lucy sensitivity denominator is unity.} the background-aware RL update is 
\begin{equation}
    \bm{\mu}^{(t+1)} = \bm{\mu}^{(t)} \circ 
    \left[ \mathbf{R}_\gamma^\tr \left( \frac{\mathbf{n}}{\mathbf{R}_\gamma\bm{\mu}^{(t)} 
    + \mathbf{b}_{\mathrm{ref}}+\epsilon}   \right)\right],
\label{eq:rl-iter}
\end{equation}
where $t=0,1,2,\dots$ is the RL iteration index, and $\circ$ and the division denote elementwise operations. The small constant $\epsilon$ is used only to avoid division by zero. In contrast to deterministic background subtraction, this update uses the ON counts directly and includes the background reference in the forward model.

The RL iteration is selected by monitoring the corresponding resolution-limited sequence
\begin{equation}
  \bm{\eta}^{(t)} = \mathbf{G}_\gamma\bm{\mu}^{(t)}.
\end{equation}
For an iteration window $w$, we define the relative deterministic change
\begin{equation}
  \Delta_\eta(t;w) = \frac{\left\|\bm{\eta}^{(t)} - \bm{\eta}^{(t-w)}\right\|_2}
{\left\|\bm{\eta}^{(t)}\right\|_2+\epsilon}.
  \label{eq:rl-delta}
\end{equation}
We estimate a Poisson-resampling noise level $\mathcal{N}_\eta(t)$ by resampling the ON counts and rerunning the RL iteration. The selected iteration is the first $t$ for which
\begin{equation}
  \frac{\Delta_\eta(t;w)}{\mathcal{N}_\eta(t)}\le\tau
  \label{eq:rl-ratio-rule}
\end{equation}
for a prescribed number of consecutive iterations. In the baseline configuration, we use $w=10$, $\tau=2.0$, 50 Poisson resamples, 10 consecutive iterations, and a maximum of 500 RL iterations. If no such iteration is found, the maximum is used.

Let $t_{\mathrm{RL}}$ denote the selected iteration. We define the emitted-space RL reference by applying the stability-floor,
\begin{equation}
  \mu_{\mathrm{RL},j}=
  \max\left\{\mu_j^{(t_{\mathrm{RL}})},\epsilon_{\mathrm{RL}}\right\}.
  \label{eq:rl-stability-floor}
\end{equation}
In the baseline configuration, we use $\epsilon_{\mathrm{RL}}=0.1$ counts. The floor prevents zero or extremely small selected RL values from becoming prior centers and is not interpreted as a physical signal contribution. Because the prior is mean-preserving around its center, such values could otherwise pull the posterior toward zero.

The corresponding resolution-limited RL reference used in the adaptive width schedule is
\begin{equation}
  \bm{\eta}_{\mathrm{RL}}=\mathbf{G}_\gamma\bm{\mu}_{\mathrm{RL}}.
\end{equation}

\subsection{Hierarchical prior on the emitted spectrum}
\label{subsec:prior}
The prior in our unfolding model must regularize the weakly identified directions of the inverse problem without imposing a degree of smoothness that biases the posterior solution. Conditioned on this prior-setting step, the latent mean layer and emitted spectrum are inferred in a fully Bayesian way. For each $\gamma$-energy bin $j$, we introduce a latent local mean $m_j$ and a bin-dependent log-scale standard deviation $\sigma_j$ for the Normal prior on $\log m_j$. The prior hierarchy is  
\begin{subequations}
\label{eq:mu-prior}
\begin{align}
  \mu_j \mid m_j &\sim \mathrm{Gamma}\left(\alpha, \beta_j = \frac{\alpha}{m_j}\right), \\
  \log m_j &\sim \mathrm{Normal}\left(\log \mu_{\mathrm{RL},j} - \frac{1}{2}\sigma_j^2, \sigma_j^2\right),
\end{align}
\end{subequations}
where $\alpha$ is the Gamma shape parameter and $\mu_{\mathrm{RL},j}$ is the stability-floored RL reference value in bin $j$, as defined in Eq.~\eqref{eq:rl-stability-floor}. This parameterization is mean-preserving relative to this reference,
\begin{equation}
  \mathbb{E}[m_j]=\mu_{\mathrm{RL},j}
  \qquad\mathrm{and}\qquad
  \mathbb{E}[\mu_j]=\mu_{\mathrm{RL},j}.
\end{equation}
The marginal squared coefficient of variation is
\begin{equation}
  \frac{\mathrm{Var}(\mu_j)}{\mathbb{E}[\mu_j]^2} = e^{\sigma_j^2} \left(1 + \frac{1}{\alpha}\right) - 1.
  \label{eq:marginal-cv}
\end{equation}
Thus, $\alpha$ and $\sigma_j$ jointly determine the relative prior dispersion around the RL-based reference level. The remaining modeling choice is the bin-dependent schedule for $\sigma_j$, which we define in the resolution-limited RL space. Let
\begin{equation}
  \bar{\eta}_{\mathrm{RL}}= \frac{1}{J} \sum_{j=1}^{J} \eta_{\mathrm{RL},j}
\end{equation}
be the mean level of the resolution-limited RL reference. We first define a shape-dependent schedule
\begin{equation}
  \sigma_{\mathrm{shape},j} = \sigma_{\min} + \frac{\sigma_{\max}-\sigma_{\min}}
{1+\eta_{\mathrm{RL},j}/\bar{\eta}_{\mathrm{RL}}}.
  \label{eq:sigma-shape}
\end{equation}
This quantity is smaller in the high-intensity regions of the RL reference and larger in weak regions. The schedule in Eq.~\eqref{eq:sigma-shape} depends only on the relative shape of the RL reference within a given spectrum, through $\eta_{\mathrm{RL},j}/\bar{\eta}_{\mathrm{RL}}$. Used by itself, this would make the largest peak in a low-statistics spectrum receive narrow prior width merely because it is large relative to the other bins in the same spectrum. This is undesirable, since a relative peak is not necessarily well constrained when the absolute number of counts is small. To make the schedule sensitive to the local counting statistics, we introduce a local activation factor
\begin{equation}
  a_j =\frac{\eta_{\mathrm{RL},j}} {\eta_{\mathrm{RL},j}+C_{\mathrm{ref}}},
\label{eq:sigma-activation}
\end{equation}
where $C_{\mathrm{ref}}>0$ is the resolution-limited count level at which the local RL-shape information is half activated. The final prior width is 
\begin{equation}
  \sigma_j = (1-a_j)\sigma_{\max} + a_j\sigma_{\mathrm{shape},j}.
\label{eq:sigma-schedule}
\end{equation}
Thus, $C_{\mathrm{ref}}$ acts as a local statistics threshold. If $\eta_{\mathrm{RL},j}\ll C_{\mathrm{ref}}$, then $a_j\approx 0$ and $\sigma_j\approx\sigma_{\max}$, even if bin $j$ is large relative to the mean of that particular spectrum. If $\eta_{\mathrm{RL},j}\gg C_{\mathrm{ref}}$, then $a_j\approx1$, and the prior follows the local shape-dependent schedule in Eq.~\eqref{eq:sigma-shape}. In this way, prominent structures are given narrower prior widths only when they have sufficient absolute resolution-limited intensity.

In the baseline configuration used throughout the main analysis, we set
\begin{equation}
  \alpha=1.0, \qquad \sigma_{\min}=1.0, \qquad \sigma_{\max}=3.0.
\end{equation}
The activation scale in Eq.~\eqref{eq:sigma-activation} is fixed at $C_{\mathrm{ref}}=100$ counts throughout the baseline analysis. With this
choice, the local RL-shape information is half activated when $\eta_{\mathrm{RL},j}=100$ counts. We treat $C_{\mathrm{ref}}$ as a fixed count-scale calibration of the adaptive width schedule, rather than as one of the prior parameters varied in the main sensitivity study.

The sensitivity analyses in Sec.~\ref{subsec:results-sensitivity} examine the posterior dependence on the Gamma shape parameter, the prior-width bounds, the RL reference, and the prior center. The uniform-width cases $\sigma_{\min}=\sigma_{\max}$ remove the $C_{\mathrm{ref}}$-dependent activation altogether and therefore test limiting versions of the adaptive schedule.

We also investigated a Fisher-information-based choice of $\sigma_j$, but found that it depended too strongly on the regularized inversion required to stabilize the ill-conditioned unfolding operator and was therefore not sufficiently robust for the prior construction.

\begin{figure}
  \centering
  \includegraphics{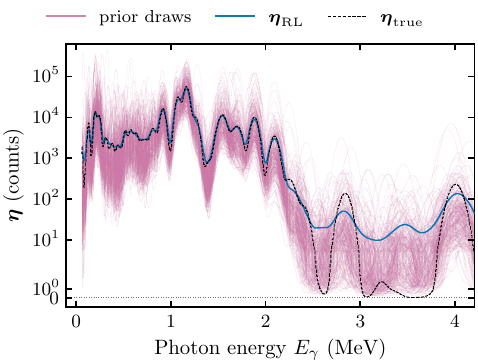}
  \caption{Prior draws in the resolution-limited space for a representative excitation-energy bin ($E_x \approx 4.0$~MeV). Thin curves show draws of $\bm{\eta}$ implied by the hierarchical prior on $\bm{\mu}$ after mapping through the $\gamma$-energy resolution operator $\mathbf{G}_\gamma$. The solid blue curve shows the RL reference estimate $\bm{\eta}_{\mathrm{RL}}$, and the black dashed curve shows the ground truth $\bm{\eta}_{\mathrm{true}}$. The spread of the prior draws visualizes the degree of prior uncertainty in $\bm{\eta}$ around the empirical scale set by $\bm{\eta}_{\mathrm{RL}}$.}
  \label{fig:prior-draws}
\end{figure}

In the present work, we use $\alpha=1$ as the baseline choice. This yields an exponential conditional prior for $\mu_j\mid m_j$, which preserves blockwise log-concavity of the conditional posterior in the $\bm{\mu}$-block while still allowing the prior mode to remain at the boundary $\mu_j=0$. Values $\alpha>1$ would introduce an additional soft repulsion from zero, but would also shift the conditional prior mode to a strictly positive value, which is less desirable in the weak-signal tail. The corresponding concavity properties are summarized in App.~\ref{app:concavity}. Figure~\ref{fig:prior-draws} illustrates representative prior draws in the resolution-limited space $\bm{\eta}$, alongside the RL reference $\bm{\eta}_{\mathrm{RL}}$ and the ground truth $\bm{\eta}_{\mathrm{true}}$. Because $\bm{\mu}_{\mathrm{RL}}$ and $\sigma_j$ are fixed by the empirical-Bayes construction before posterior sampling, uncertainty from this prior-setting step is not propagated directly. Instead, its practical impact is assessed through the sensitivity analyses in Sec.~\ref{subsec:results-sensitivity}. From a computational perspective, conditioning on this fixed prior construction leads to a posterior that is considerably more tractable than the fully Bayesian alternatives considered during development, while avoiding the bias introduced by stronger smoothness priors.

\subsection{Stochastic background model}
\label{subsec:bgmodel}

In the Oslo method, the detected counts consist of both signal and background contributions. The objective of the background model is to separate these components while formally propagating the statistical uncertainty inherent in the background estimation. Each $E_\gamma$ bin $j$ is assigned a latent background expectation $b_j$, which is modeled independently across the spectrum using a Gamma prior
\begin{equation}
  b_j \sim \Gam(a_0, b_0).
  \label{eq:prior-lambda}
\end{equation}
We use the rate parameterization of the Gamma distribution, so that $\mathbb{E}[b_j]=a_0/b_0$. In the baseline analysis, we set $a_0=1.0$ and choose 
\begin{equation}
  b_0 = \frac{a_0}{\bar{n}_{\mathrm{off}}},
  \qquad
  \bar{n}_{\mathrm{off}} = \frac{1}{J}\sum_{j=1}^{J} n_{\mathrm{off},j}.
\end{equation}
This gives a broad empirical prior centered on the average OFF-count level, while the binwise background levels remain primarily constrained by the observed OFF counts.

The model assumes that the background measurement taken in the absence of signal (OFF run) is statistically stationary and representative of the background contributing to the total counts in the signal (ON) runs. Under this assumption, there is no systematic bias between the two configurations, and the binwise information about $b_j$ comes from the Poisson counting statistics of the OFF data. By treating $b_j$ as a latent stochastic variable, the framework propagates the background uncertainty into the final spectral uncertainty of $\bm{\eta}$. This Bayesian formulation replaces deterministic background subtraction with a joint probabilistic inference, enforcing physical non-negativity even in regimes where the background intensity is comparable to the total observed counts.

\subsection{Posterior sampling with NUTS} 
\label{subsec:nuts} 

\begin{figure*}
\centering 
\includegraphics{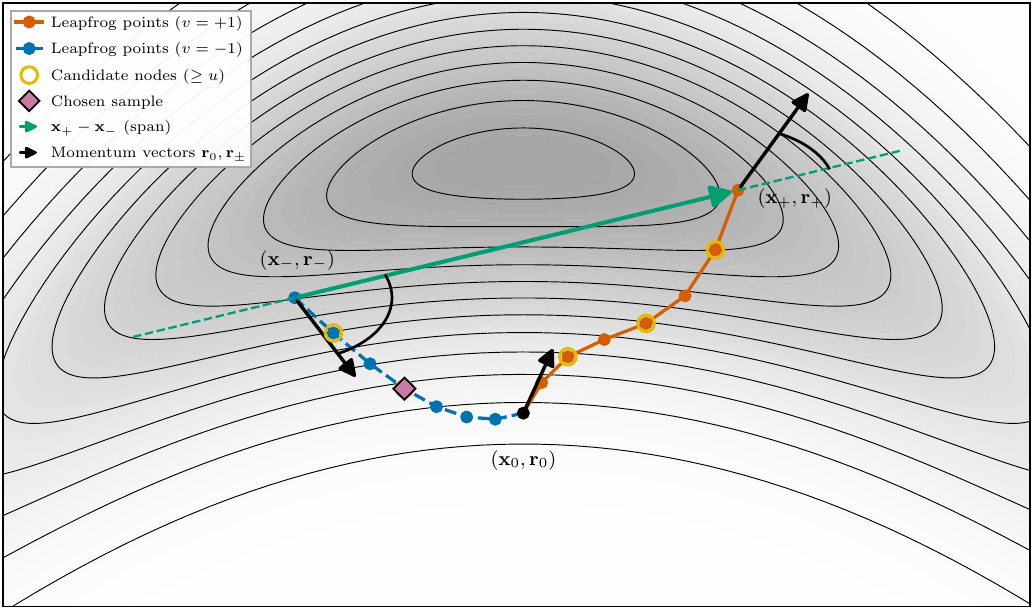} 
\caption{Schematic illustration of one No-U-Turn Sampler transition on a two-dimensional target density. Starting from $(\mathbf{x}_0,\mathbf{r}_0)$, leapfrog states are generated forward $(v=+1)$ and backward $(v=-1)$ in fictitious time, producing endpoint states $(\mathbf{x}_+,\mathbf{r}_+)$ and $(\mathbf{x}_-,\mathbf{r}_-)$. The green arrow indicates the span $\mathbf{x}_+ - \mathbf{x}_-$ between the two endpoints, while the black arrows indicate the momentum directions used in the No-U-Turn criterion. Yellow circles mark states satisfying the slice criterion, and the purple diamond indicates the candidate selected as the next Markov-chain state.}
\label{fig:NUTS} 
\end{figure*}

We sample from the posterior using NUTS as implemented in \texttt{PyMC}, following the Hamiltonian Monte Carlo (HMC) framework of Neal~\cite{neal2011mcmc} and the NUTS construction of Hoffman and Gelman~\cite{hoffman2014nuts}. NUTS is an adaptive HMC variant that extends the (unconstrained) parameter vector $\mathbf{x}$ with a fictitious momentum vector $\mathbf{r}$ of the same dimension. The Hamiltonian
\begin{equation}
  \mathcal{H}(\mathbf{x},\mathbf{r})
  = -\log \pi(\mathbf{x} \mid \mathbf{n}, \mathbf{n}_{\mathrm{off}})
    + \frac{1}{2} \mathbf{r}^\tr \bm{\mathcal{M}}^{-1} \mathbf{r},
  \label{eq:hamiltonian}
\end{equation}
combines the negative log posterior with a quadratic kinetic term defined by the mass matrix $\bm{\mathcal{M}}$. Leapfrog integration uses the gradient $\nabla_{\mathbf{x}}\log \pi(\mathbf{x}\mid\mathbf{n},\mathbf{n}_{\mathrm{off}})$ to update $\mathbf{r}$
and then moves $\mathbf{x}$ along approximate Hamiltonian trajectories, proposing distant states with high acceptance probability.

A major practical difficulty in HMC is choosing the trajectory length, or equivalently the number of leapfrog steps. NUTS avoids this manual tuning by recursively building a binary tree of leapfrog states in fictitious time. At each doubling step, a new branch is generated either forward $(v=+1)$ or backward $(v=-1)$ from the current tree. At the beginning of each iteration, a new momentum $\mathbf{r}_0 \sim \Normal(\mathbf{0},\bm{\mathcal{M}})$ is drawn, and a slice variable
\begin{equation}
  u \sim \mathrm{Uniform}\bigl(0, \exp[-\mathcal{H}(\mathbf{x}_0, \mathbf{r}_0)]\bigr),
  \label{eq:slice}
\end{equation}
defines a threshold in the joint density over $(\mathbf{x},\mathbf{r})$. Only states with $\exp[-\mathcal{H}(\mathbf{x},\mathbf{r})] \ge u$ are eligible as candidates for the next draw. Tree expansion proceeds by recursive doubling until either the No-U-Turn criterion is met or the configured maximum tree depth is reached. If $\mathbf{x}_+$ and $\mathbf{x}_-$ denote the forward and backward endpoints of the current tree, with corresponding momenta $\mathbf{r}_+$ and $\mathbf{r}_-$, the No-U-Turn criterion can be written as
\begin{equation}
  (\mathbf{x}_+ - \mathbf{x}_-)^{\tr}\mathbf{r}_+ < 0
  \quad\mathrm{or}\quad
  (\mathbf{x}_+ - \mathbf{x}_-)^{\tr}\mathbf{r}_- < 0.
  \label{eq:nuts-uturn}
\end{equation}
Thus, expansion stops once either endpoint momentum has a negative projection onto the span $\mathbf{x}_+ - \mathbf{x}_-$, indicating that the simulated trajectory has begun to turn back on itself. Figure~\ref{fig:NUTS} illustrates this construction schematically. At the end of the recursion, one of the slice-eligible states visited in the tree is selected as the next Markov-chain state (following Algorithm~6 in Ref.~\cite{hoffman2014nuts}) in a way that preserves the posterior distribution as the stationary distribution of the chain.

During the warm-up phase, NUTS adaptively tunes the leapfrog step size $\varepsilon$ and the mass matrix $\bm{\mathcal{M}}$ to better match the posterior geometry, typically improving sampling efficiency compared to fixed, manually chosen values. The concrete backend choice and tuning configuration used in this work are provided in
Sec.~\ref{subsec:sampler-config}, and diagnostic criteria are described in Sec.~\ref{subsec:diagnostics}.

\subsection{Data informativeness and log-parameterization}
\label{subsec:data-info}
The positive variables in the model are represented internally on an unconstrained scale for NUTS sampling. The emitted spectrum and latent background expectation are represented through logarithmic coordinates,
\begin{equation}
  \bm{\phi} = \log\bm{\mu}, \qquad 
  \bm{\psi} = \log\mathbf{b}.
\end{equation}
The latent mean layer in Eq.~\eqref{eq:mu-prior} is implemented in a non-centered form. Specifically, we write
\begin{equation}
  \log m_j =\log\mu_{\mathrm{RL},j}-\frac{1}{2}\sigma_j^2+\sigma_j z_{m,j},
  \label{eq:noncentered-logm}
\end{equation}
where $z_{m,j}\sim \Normal(0,1)$. This is equivalent to the lognormal layer in Eq.~\eqref{eq:mu-prior}, but it uses standardized latent variables $\mathbf{z}_m$ as the sampled coordinates. When the background is included as a latent variable, the unconstrained coordinates sampled by NUTS are therefore $(\bm{\phi},\mathbf{z}_m,\bm{\psi})$, not the positive variables $(\bm{\mu},\mathbf{m},\mathbf{b})$ themselves.

The emitted spectrum enters the likelihood through the folded signal expectation $\bm{\nu}=\mathbf{R}_\gamma\bm{\mu}$, whereas the latent mean coordinates $\mathbf{z}_m$ influence the posterior through the hierarchical prior coupling between $\mathbf{m}$ and $\bm{\mu}$. The relevant question for data informativeness is therefore how a move in the log-emitted-spectrum coordinates changes the expected detected signal. Consider a small perturbation
\begin{equation}
  \bm{\phi}\mapsto \bm{\phi}+\epsilon\mathbf{h},
\end{equation}
where $\mathbf{h}$ is an arbitrary direction and $\epsilon$ is small. Since $\mu_j=\exp(\phi_j)$, the emitted spectrum changes to first order as 
\begin{equation}
  \delta\mu_j = \epsilon \mu_j h_j + \mathcal{O}(\epsilon^2).
\end{equation}
The corresponding first-order change in the folded signal expectation is 
\begin{equation}
  \delta\nu_i = \epsilon\sum_j R_{\gamma,ij}\mu_j h_j + \mathcal{O}(\epsilon^2).
  \label{eq:delta-nu-log}
\end{equation}
Equation~\eqref{eq:delta-nu-log} shows why the absolute count scale matters for the posterior geometry. A finite perturbation in $\phi_j$ produces an absolute change in $\mu_j$ proportional to the current intensity $\mu_j$. Accordingly, perturbations in low-intensity regions can produce only small changes in the expected detected signal, even when the perturbation is non-negligible on the log scale. Such directions tend to have weak likelihood curvature and are therefore more strongly influenced by the prior. In contrast, directions that produce significant changes in the folded expectation $\bm{\nu}$ are more strongly data-driven. 

This argument should not be interpreted bin by bin. The response matrix couples the emitted-energy bins, so the weakly and strongly informed directions are generally linear combinations of bins rather than individual coordinates. For fixed background expectation, the local Fisher information in the log-emitted coordinates is 
\begin{equation}
  \bm{\mathcal I}_{\phi}(\bm{\phi})=\mathrm{diag}(\bm{\mu})\mathbf{R}_\gamma^\tr
  \mathrm{diag}(\bm{\lambda})^{-1}\mathbf{R}_\gamma\mathrm{diag}(\bm{\mu}),
  \label{eq:fisher-log-coordinates}
\end{equation}
where $\bm{\lambda}=\mathbf{R}_\gamma\bm{\mu}+\mathbf{b}$.
This expression makes explicit that the likelihood curvature depends both on the response operator and on the local intensity scale. Directions that are strongly attenuated by $\mathbf{R}_\gamma$, or that live in regions where the absolute intensity is small, carry less likelihood information. This separation between data-dominated and prior-dominated directions motivates the prior-width schedule in Sec.~\ref{subsec:prior}. The prior is broader in weak-count regions and narrower where the resolution-limited RL reference indicates substantial local count level.

\subsection{Sampler backend and tuning configuration}
\label{subsec:sampler-config}

The statistical model is specified as a \texttt{PyMC} model. This model can be sampled with the native \texttt{PyMC} NUTS implementation or with external NUTS backends available through the \texttt{PyMC} interface. The main results reported in this article use the native \texttt{PyMC} backend. To check that the implementation was not tied to a single sampler backend, we benchmarked the same model with \texttt{NumPyro}~\cite{phan2019numpyro} and \texttt{BlackJAX}~\cite{cabezas2024blackjax}. We also evaluated \texttt{Nutpie}~\cite{seyboldt2026nutpie}, but under the fixed configuration used here it did not satisfy the diagnostic stability criterion and is therefore omitted from the timing table. This omission should be interpreted only for this particular model, initialization, target-acceptance schedule, and backend configuration, not as a general assessment of the sampler backend. The matched backend comparison is reported in Table~\ref{tab:benchmarks} in App.~\ref{app:benchmarks}. Overall wall-clock performance depends on the excitation energy, count statistics, number of active $E_\gamma$ bins, sampler configuration, and backend.

Each reported inference run uses four chains. For each chain, we run 2000 warm-up iterations followed by 2000 posterior draws. The maximum tree depth is set to 13. During warm-up, NUTS adapts the leapfrog step size and a diagonal mass matrix in the unconstrained log-parameterization described in Sec.~\ref{subsec:data-info}. Dense mass-matrix adaptation was also tested during method development, but it did not improve the sampling efficiency for this model and increased computational cost.

The NUTS target acceptance probability is selected according to the excitation-energy range. We use a conservative value for low excitation-energy bins, where the posterior geometry is typically more challenging, and a lower value for high excitation-energy bins. In the synthetic-data studies reported here, the target acceptance is 0.99 for $E_x<3.0$~MeV, 0.95 for $3.0\le E_x<6.0$~MeV, and 0.90 for higher excitation energies. Divergent transitions and maximum-tree-depth saturation are monitored as described in Sec.~\ref{subsec:diagnostics}.

\subsection{Diagnostics and convergence}
\label{subsec:diagnostics}

\begin{figure*}
  \centering
  \includegraphics{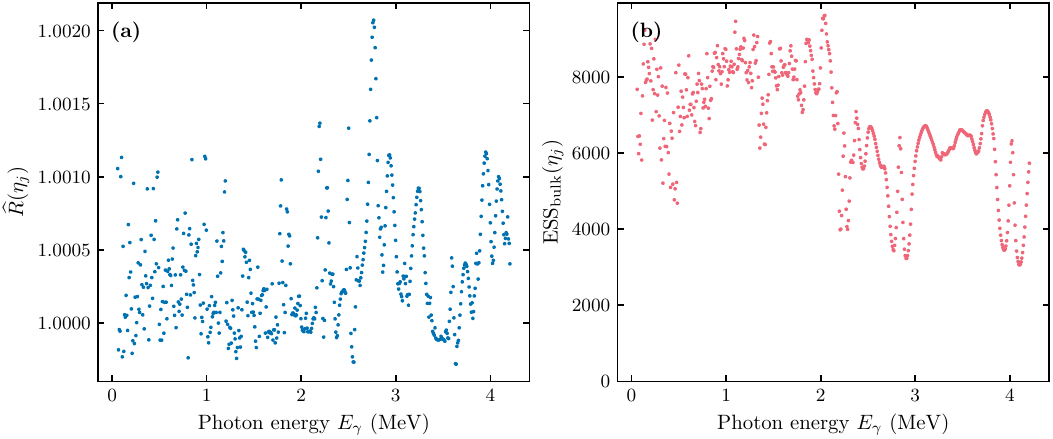}%
  \caption{Sampling diagnostics per $\gamma$-energy bin for the inferred resolution-limited spectrum $\bm{\eta}$ at $E_x \approx 4.0$~MeV. (a) Rank-normalized split $\widehat{R}(\eta_j)$. (b) Bulk effective sample size $\mathrm{ESS}_{\mathrm{bulk}}(\eta_j)$. Diagnostics are computed separately for each $E_\gamma$ bin using the posterior draws. Values of $\widehat{R}$ close to one and large $\mathrm{ESS}_{\mathrm{bulk}}$ indicate good mixing and small Monte Carlo error relative to the posterior uncertainty.}
  \label{fig:sampling-diagnostics}
\end{figure*}

Convergence and sampling efficiency are assessed using the rank-normalized split $\widehat{R}$ statistic and the bulk effective sample size $\mathrm{ESS}_{\mathrm{bulk}}$, as implemented in the \texttt{ArviZ} library~\cite{arviz2019, vehtari2021rank}. 
The $\widehat{R}$ statistic monitors the consistency between independent Markov chains, where values near 1.00 are consistent with convergence of the chains to a common stationary distribution. The $\mathrm{ESS}_{\mathrm{bulk}}$ provides an estimate of the number of independent samples contained in the autocorrelated posterior draws, serving as a measure of the sampling precision.

Since the primary reported quantity is the resolution-limited spectrum $\bm{\eta}$ for each excitation-energy bin $k$ (see Sec.~\ref{subsec:bayes}), we compute diagnostics per $\gamma$-energy bin $j$ on $\bm{\eta}$ rather than on the latent emitted intensity $\bm{\mu}$. To evaluate the overall convergence of the spectrum for a given excitation-energy bin, we report the worst-case 
$\widehat{R}$ across all $\gamma$-energy bins, $\widehat{R}_{\max} = \max_j \widehat{R}(\eta_j)$. Similarly, we characterize the sampling efficiency by the minimum $\mathrm{ESS}_{\mathrm{bulk}}$ across all $\gamma$-energy bins. The minimum $\mathrm{ESS}_{\mathrm{bulk}}$ is particularly informative, as the unfolded spectrum typically contains low-count regions at high $E_\gamma$ where autocorrelation tends to be higher.

Figure~\ref{fig:sampling-diagnostics} shows a representative diagnostic profile for the spectrum at an excitation energy of $E_x \approx 4.0$~MeV. Values of $\widehat{R} < 1.01$ across all $\gamma$-energy bins, combined with a large $\mathrm{ESS}_{\mathrm{bulk}}$, indicate that the chains have mixed well and that the Monte Carlo error in the posterior sampling is small compared with the posterior spread. In addition to $\widehat{R}$ and ESS, we monitor two NUTS-specific diagnostics:
\begin{itemize}
    \item \textbf{Divergent transitions} (\textit{Div.}): These indicate numerical instabilities in the Hamiltonian integration, signaling regions where the sampler may fail to reliably explore the posterior.
    \item \textbf{Maximum tree-depth saturation} (\textit{MTD}\%): This represents the fraction of posterior samples that reach the configured maximum tree depth. A non-zero \textit{MTD}\% indicates that the trajectory-length cap is limiting the exploration before the No-U-Turn criterion is met, which can reduce sampling efficiency and may indicate a complex posterior geometry.
\end{itemize}

\subsection{Posterior uncertainty via global rank envelopes}
\label{subsec:rank-envelope}
Each posterior draw of $\bm{\eta}$ is a full spectrum over $E_\gamma$. Therefore, the uncertainty summary should respect the strong correlation across photon energies. We use simultaneous uncertainty bands based on the global rank-envelope method, as implemented in the GET framework~\cite{myllymaki2017global,myllymaki2024get}. We fix an $E_x$ bin and write the $N$ posterior draws as
\begin{equation}
  \bm{\eta}^{(s)} = \left(\eta^{(s)}_1, \dots, \eta^{(s)}_J \right), \quad s=1, \dots, N,
\end{equation}
defined on a common $E_\gamma$ grid $j=1,\dots,J$.
For each $\gamma$-energy bin $j$, we compute the pointwise ranks
\begin{equation}
  r_{s,j} = \mathrm{rank} \left( \eta^{(s)}_j \,;\, \eta^{(1)}_j, \dots, \eta^{(N)}_j \right) \in \{1, \dots, N\},
\end{equation}
where rank $1$ corresponds to the smallest value and rank $N$ to the largest. Ties are handled by mid-ranks. To obtain a two-sided notion of extremeness (low or high), we convert to two-sided ranks
\begin{equation}
  R_{s,j} = \min \{r_{s,j}, \, N + 1 - r_{s,j}\}.
\end{equation}
A small $R_{s,j}$ indicates that draw $s$ is extreme at bin $j$. To rank entire spectral curves by their joint extremeness over all $E_\gamma$ bins, we use the extreme rank length (ERL) ordering. For each draw $s$, we sort its two-sided ranks in ascending order to form the vector $\mathbf{R}^{\uparrow}_s = \mathrm{sort}(R_{s,1}, \dots, R_{s,J})$. Curves are compared lexicographically: if two curves share the same minimum rank, the tie is broken by comparing their next smallest rank, and so on. A curve is deemed more extreme if the first differing entry in its sorted vector is the smaller of the two. This ERL ordering provides a stable tie-breaking refinement compared to using only the minimum rank across bins. 

Given an envelope mass level $m=0.95$, we retain the $N_m = \lceil mN \rceil$ least extreme draws according to the ERL ordering. The global envelope band is then defined pointwise as the minimum and maximum over this kept subset $\mathcal{K}$:
\begin{equation}
  L_j = \min_{s \in \mathcal{K}} \eta^{(s)}_j, \qquad U_j = \max_{s \in \mathcal{K}} \eta^{(s)}_j.
\end{equation}
Because draws are discarded based on their extremeness across all $j$, the resulting band $[L_j, U_j]$ is simultaneous over the $E_\gamma$ axis. In the figures presented in Sec.~\ref{sec:results}, we plot the posterior mean $\bar{\eta}_j$ as a central summary. The global envelope itself is constructed from the posterior draws and does not require the marginal posteriors to be symmetric.

\subsection{Prior and posterior predictive checks}
\label{subsec:ppc}

We assess model adequacy through empirical prior predictive and posterior predictive checks in the observed count space, conditional on the data-driven prior-setting step described above. In the prior predictive checks, we sample the emitted intensity and background from their respective priors,
\begin{equation}
  \bm{\mu} \sim \pi(\bm{\mu}), \qquad \mathbf{b} \sim \pi(\mathbf{b}),
\end{equation}
calculate the expected detected signal $\bm{\nu} = \mathbf{G}_\gamma \mathbf{D} \bm{\mu}$, and simulate replicated datasets:
\begin{equation}
  \tilde{\mathbf{n}}_{\mathrm{off}} \sim \Pois(\mathbf{b}), \qquad \tilde{\mathbf{n}} \sim \Pois(\bm{\nu} + \mathbf{b}).
\end{equation}
By comparing the prior predictive bands for $\tilde{\mathbf{n}}$ and $\tilde{\mathbf{n}}_{\mathrm{off}}$ to the observed counts, we check whether the hierarchical prior places sufficient probability mass in the physically relevant regions of the data space.

In the posterior predictive checks, we repeat this procedure using draws from the joint posterior $\pi(\bm{\mu}, \mathbf{b} \mid \mathbf{n}, \mathbf{n}_{\mathrm{off}})$. This generates posterior predictive envelopes for the total counts and background, which we compare to the observed $\mathbf{n}$ and $\mathbf{n}_{\mathrm{off}}$. Furthermore, we compare the posterior bands for the folded signal expectation $\bm{\nu}$ with the truth $\bm{\nu}_{\mathrm{true}}$. This checks whether the model is consistent with both the observed count data and the underlying folded-signal expectation in the synthetic test.

\subsection{Sensitivity and robustness analysis}
\label{subsec:sensitivity-robustness}

We quantify the stability of the unfolded spectra under alternative modeling choices using a scaled-deviation statistic. Let $\bm{\eta}_{\mathrm{base}}^{(l)}$ and $\bm{\eta}_{\mathrm{alt}}^{(l)}$ ($l=1, \dots, L$) denote matched sets of posterior draws of the resolution-limited spectrum for a fixed excitation-energy bin under the baseline and alternative configurations, respectively. First, we compute the baseline 95\% global rank-envelope $[L_j, U_j]$ for the baseline draws using the procedure described in Sec.~\ref{subsec:rank-envelope}. We then define the binwise scale factor $w_{j,1/2}$ as the envelope half-width
\begin{equation}
  w_{j,1/2} = \max \left\{ \frac{U_j - L_j}{2}, \, \epsilon \right\},
\end{equation}
where $\epsilon$ is a small numerical floor for stability. To assess the impact of the modeling change, we form the scaled deviations by pairing the independent baseline and alternative draws. For each draw $l \in \{1, \dots, L\}$, we define the scaled deviation
\begin{equation}
  S_j^{(l)} = \frac{\eta^{\mathrm{alt},(l)}_j - \eta^{\mathrm{base},(l)}_j}{w_{j,1/2}}.
  \label{eq:scaled-deviation-alt}
\end{equation}
Values of $S_j^{(l)}=\pm1.0$ correspond to deviations of one baseline half-width, $w_{j,1/2}$. Larger values indicate that the displacement exceeds this reference scale. To establish a baseline reference for the level of variation expected from posterior sampling alone under the baseline model, we construct a null distribution by replacing the alternative draws with an independent permutation of the baseline draws.

In validation figures comparing posterior draws to the known synthetic truth, we employ an analogous scaled deviation
\begin{equation}
  S_j^{(l)}=\frac{\eta_j^{(l)}-\eta_{\mathrm{true},j}}{w_{j,1/2}}.
\label{eq:scaled-deviation-truth}
\end{equation}
The physical significance of a scaled deviation must always be interpreted together with the absolute count scale. This is especially important in low-statistics regions close to the non-negativity boundary. There, the baseline envelope half-width can be small in absolute count units, so minor changes in $\eta_j$ may correspond to large values of $S_j^{(l)}$. Hence, the scaled deviation is used as a relative diagnostic, while the targeted low-statistics prior-dependence check in Sec.~\ref{subsec:results-lowstat-prior-dependence} is interpreted directly on the absolute $\bm{\eta}$ scale.

\section{Results}
\label{sec:results}

We validate the Bayesian unfolding model on synthetic data for which the ground-truth spectra are known. First, we examine a representative baseline case at $E_x \approx 4.0$~MeV, focusing on the structure of the empirical-Bayes prior, the posterior recovery of the resolution-limited spectrum $\bm{\eta}$, prior-to-posterior contraction, and predictive checks in the observed-count space. Using the same high-statistics case, we then quantify sensitivity to the main empirical-Bayes hyperparameters and robustness to selected modeling assumptions. Keeping the count level and spectral structure fixed allows the individual impact of each variation to be isolated. Next, we test the method under low-statistics conditions using representative $E_x\approx2.5$~MeV and $E_x\approx9.5$~MeV spectra. Since the low-statistics regime is where prior specifications are most likely to matter, we also perform a targeted prior-dependence check for the $E_x\approx9.5$~MeV spectrum. Finally, we compare the Bayesian posterior results with the frequentist RMLE framework recently developed for unfolding $\gamma$-ray spectra in nuclear-physics applications~\cite{lima2025}.

\subsection{\boldmath Bayesian unfolding at \texorpdfstring{$E_x \approx 4.0$}{Ex approx 4.0}~MeV}
\label{subsec:baseline-prior-4MeV}

\begin{figure}
  \centering
  \includegraphics{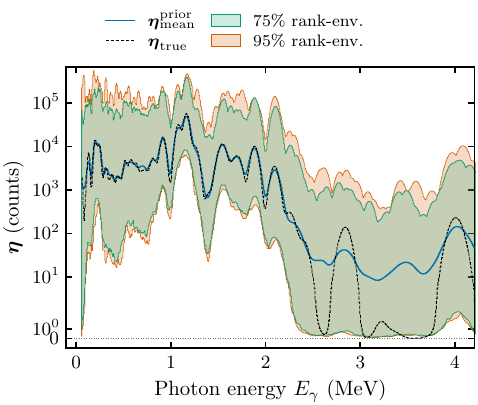}
  \caption{Empirical-Bayes prior in the resolution-limited space for the $E_x \approx 4.0$~MeV excitation-energy bin. The solid blue curve shows the prior mean, and the black dashed curve shows the ground truth $\bm{\eta}_{\mathrm{true}}$. The shaded regions show the simultaneous 75\% and 95\% global rank-envelope bands. The prior is centered on the stability-floored RL reference and uses the adaptive $\sigma_j$ schedule from Eq.~\eqref{eq:sigma-schedule}, with $(\sigma_{\min},\sigma_{\max})=(1.0,3.0)$.}
  \label{fig:baseline-prior}
\end{figure}

We first examine the prior induced by the empirical-Bayes construction. Figure~\ref{fig:baseline-prior} illustrates the prior distribution after mapping the emitted spectrum through the detector-resolution operator, $\bm{\eta}=\mathbf{G}_\gamma\bm{\mu}$. The prior mean follows the scale set by the RL reference. The purpose of the prior is to provide a positive, physically plausible scale while remaining broad enough to cover alternative spectra that are compatible with the ill-conditioned forward problem.

The nested 75\% and 95\% rank-envelope bands show that the prior uncertainty is substantial, especially in weakly constrained regions. This behavior reflects the adaptive width schedule in Eq.~\eqref{eq:sigma-schedule}. Local RL-shape information is allowed to narrow the prior only where the resolution-limited count level is sufficiently large. In the high-$E_\gamma$ tail, where the absolute count level is small, the prior width stays close to $\sigma_{\max}$.

\begin{figure*}
  \centering
  \includegraphics{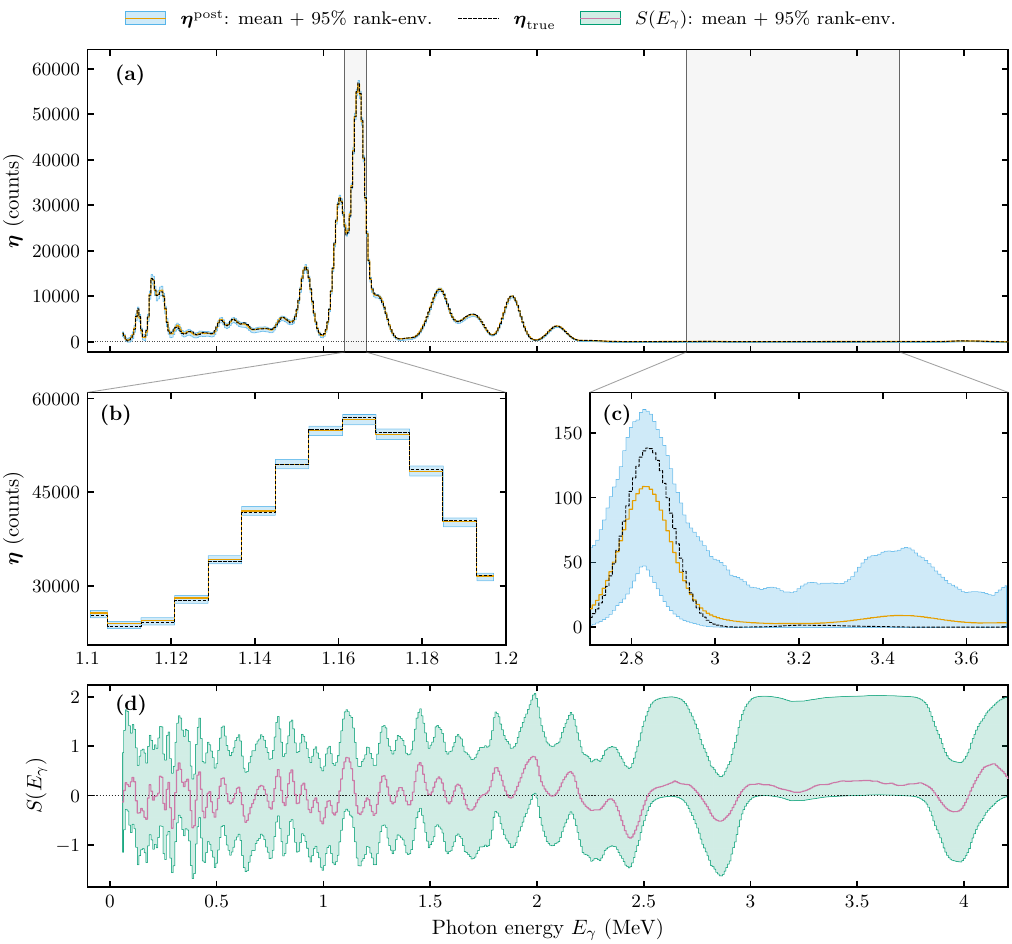}
  \caption{Bayesian posterior in the resolution-limited space at $E_x \approx 4.0$~MeV. (a) Full spectral range showing the posterior mean, 95\% global rank-envelope, and ground truth $\bm{\eta}_{\mathrm{true}}$ as a black dashed curve. (b), (c) Local zoom panels around representative spectral structures. (d) Scaled deviation $S(E_\gamma)=(\eta-\eta_{\mathrm{true}})/w_{1/2}$, where $w_{1/2}$ is the posterior-envelope half-width. The dotted horizontal line marks zero deviation from the truth.}
  \label{fig:baseline-post}
\end{figure*}

The corresponding posterior distribution is shown in Fig.~\ref{fig:baseline-post}. On the full-spectrum scale in panel (a), the posterior mean is almost indistinguishable from the ground truth. The zoom panels (b) and (c) show that the main spectral structures are recovered locally within the posterior uncertainty. The high-$E_\gamma$ region illustrates an important boundary effect that also appears in the low-statistics examples below. In bins where the resolution-limited intensity is close to zero, the non-negative prior on $\bm{\mu}$, together with the non-negative resolution operator $\mathbf{G}_\gamma$, imposes a lower boundary on posterior draws of $\bm{\eta}=\mathbf{G}_\gamma\bm{\mu}$. The lower edge of the posterior band is then constrained near zero. At the same time, the likelihood is weak in this part of the spectrum, so the posterior does not collapse onto the boundary. Instead, the uncertainty becomes asymmetric. The posterior mean is pulled toward the near-zero region, while the upper part of the envelope still allows a small positive intensity range. This is the expected geometry of a non-negative Poisson inverse problem in a weakly constrained region, not an artifact of deterministic clipping.

Fig.~\ref{fig:baseline-post}(d) shows the comparison after normalizing deviations from the truth by the 95\% posterior-envelope half-width. In the signal-rich region, the posterior mean is close to the truth on the absolute scale, and the posterior band is narrow. The mean scaled deviation shows alternating positive and negative values, remaining below one posterior half-width and mostly substantially smaller. Away from the non-negativity boundary, the scaled envelope is approximately symmetric around the mean. Note that the scaled envelope with respect to the true spectrum has a fixed range of two by construction, so a displacement of the mean shifts the envelope accordingly relative to the zero-deviation line. In the high-$E_\gamma$ tail, where the truth approaches the boundary, the envelope becomes one-sided for the reasons described above. In total, the scaled-deviation panel shows that the 95\% rank-envelope contains the truth in the signal-rich region, while demonstrating the expected asymmetric uncertainty structure in the weakly constrained high-$E_\gamma$ region.

\begin{figure}
  \centering
  \includegraphics{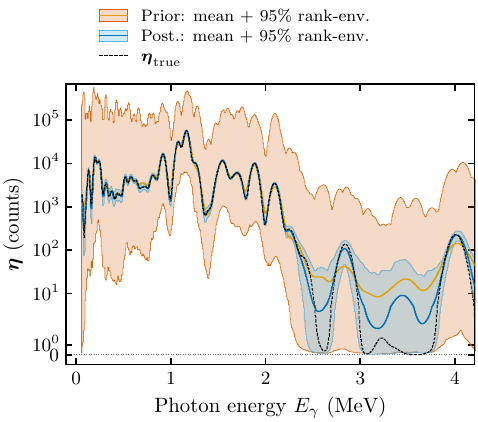}
  \caption{Prior-to-posterior contraction in the resolution-limited space at $E_x \approx 4.0$~MeV. The prior and posterior are each summarized by their mean and 95\% rank-envelope. The black dashed curve shows $\bm{\eta}_{\mathrm{true}}$.}
  \label{fig:prior-vs-post}
\end{figure}

The prior-to-posterior contraction is shown in Fig.~\ref{fig:prior-vs-post}. The contraction is strongest in the high-count spectral structures, where the data are highly informative about the resolution-limited spectrum. In these regions, the posterior band is much narrower than that of the prior. In the high-$E_\gamma$ tail, the contraction is weaker because the expected signal is low and the likelihood curvature is small. The posterior therefore balances data-driven constraints in the informative regions with broader prior-regularized uncertainty in the weakly identified tail.

\begin{figure*}
  \centering
  \includegraphics{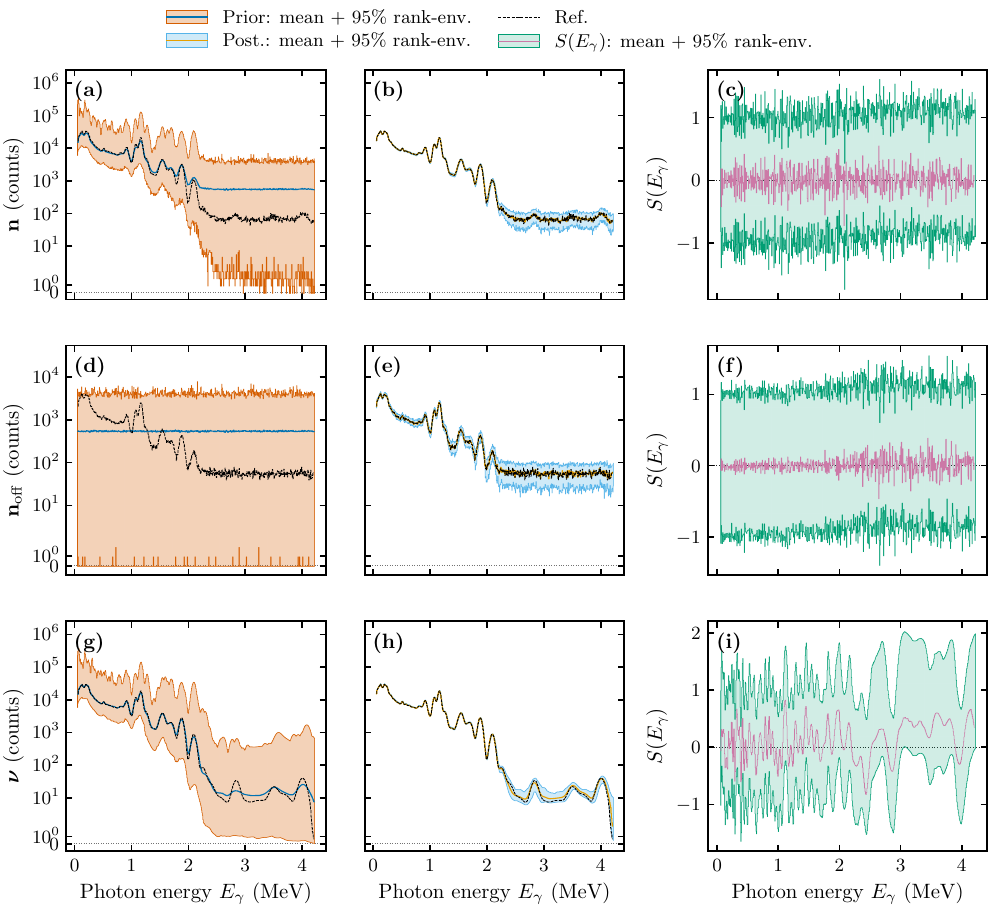}
  \caption{Prior and posterior predictive checks for the $E_x \approx 4.0$~MeV excitation-energy bin. Columns show the prior predictive summaries, posterior predictive summaries, and scaled deviations. Rows show (a)--(c) observed ON counts $\mathbf{n}$, (d)--(f) OFF/background counts $\mathbf{n}_{\mathrm{off}}$, and (g)--(i) the folded signal expectation $\bm{\nu}$. Dashed curves indicate the observed counts for the ON/OFF rows and the synthetic truth $\bm{\nu}_{\mathrm{true}}$ for the folded-signal row. The right column shows scaled deviations relative to the corresponding reference curve.}
  \label{fig:model-checks}
\end{figure*}

We evaluate model adequacy through prior and posterior predictive checks in Fig.~\ref{fig:model-checks}. The top row displays the predictive distributions for the observed ON counts $\mathbf{n}$, the middle row shows predictive distributions for the OFF counts $\mathbf{n}_{\mathrm{off}}$, and the bottom row compares the folded signal expectation $\bm{\nu}$ with its synthetic truth. The prior predictive bands are sufficiently broad to encompass the observed count levels and the folded signal scale, indicating that the empirical-Bayes prior is not overly restrictive. After conditioning on the data, the posterior predictive bands contract strongly. The posterior predictive means follow the observed ON and OFF counts, and the folded signal expectation aligns with $\bm{\nu}_{\mathrm{true}}$. Furthermore, the scaled deviation panels exhibit no systematic displacement from the respective reference curves. This is consistent with the model simultaneously describing the total observed counts, the background measurement, and the latent folded signal for this high-statistics case.

\subsection{Sensitivity and robustness}
\label{subsec:results-sensitivity}
\begin{figure}
  \centering
\includegraphics{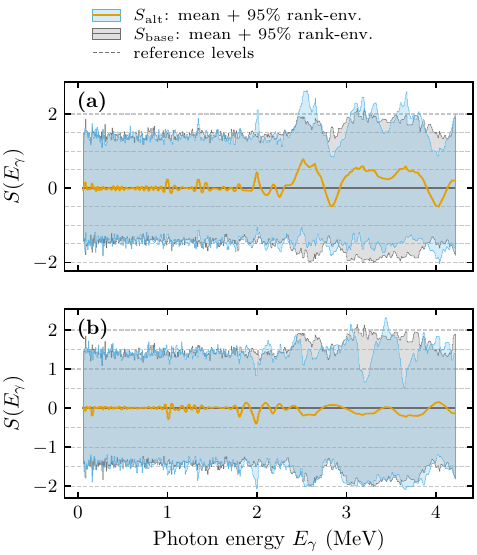}
  \caption{Sensitivity to the RL reference iteration used in the empirical-Bayes prior construction. The baseline prior uses the automatic semi-convergence rule, which selected $t_{\mathrm{RL}}^{\mathrm{auto}}=58$ for the $E_x\approx4.0$~MeV spectrum. Panels show alternatives with (a) $t_{\mathrm{RL}}=5$ and (b) $t_{\mathrm{RL}}=500$. The colored band and curve show $S_{\mathrm{alt}}$, while the gray band shows the baseline reference $S_{\mathrm{base}}$. Both alternatives remain small on the scale of the baseline posterior uncertainty, with the under-iterated case producing the larger visible shift.}
  \label{fig:rl-sensitivity}
\end{figure}

\begin{figure*}
  \centering
  \includegraphics{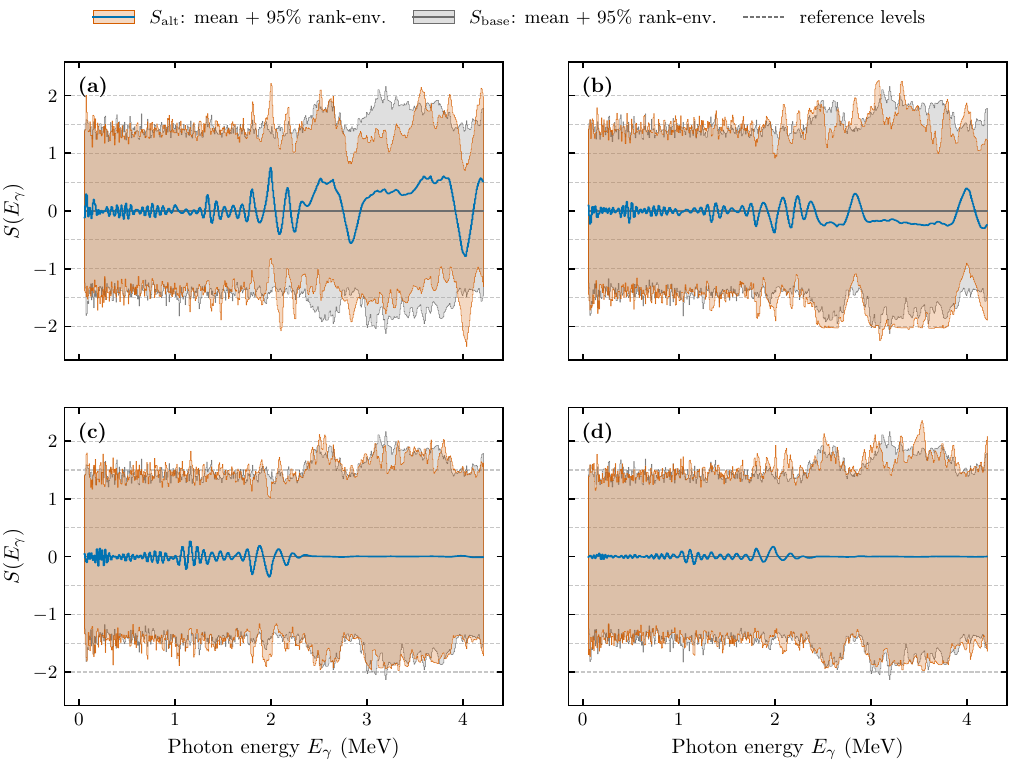}
  \caption{Sensitivity to the adaptive prior-width bounds $(\sigma_{\min}, \sigma_{\max})$, compared with the baseline $(1.0, 3.0)$. Panels show (a) $(1.0, 1.0)$, (b) $(1.0, 5.0)$, (c) $(3.0, 3.0)$, and (d) $(0.3, 3.0)$. The most visible shift occurs for the uniform $\sigma_j=1.0$ prior-width schedule in panel (a), which reduces the broad tail behavior of the baseline schedule.}
  \label{fig:sigma-sensitivity}
\end{figure*}

\begin{figure*}
  \centering
  \includegraphics{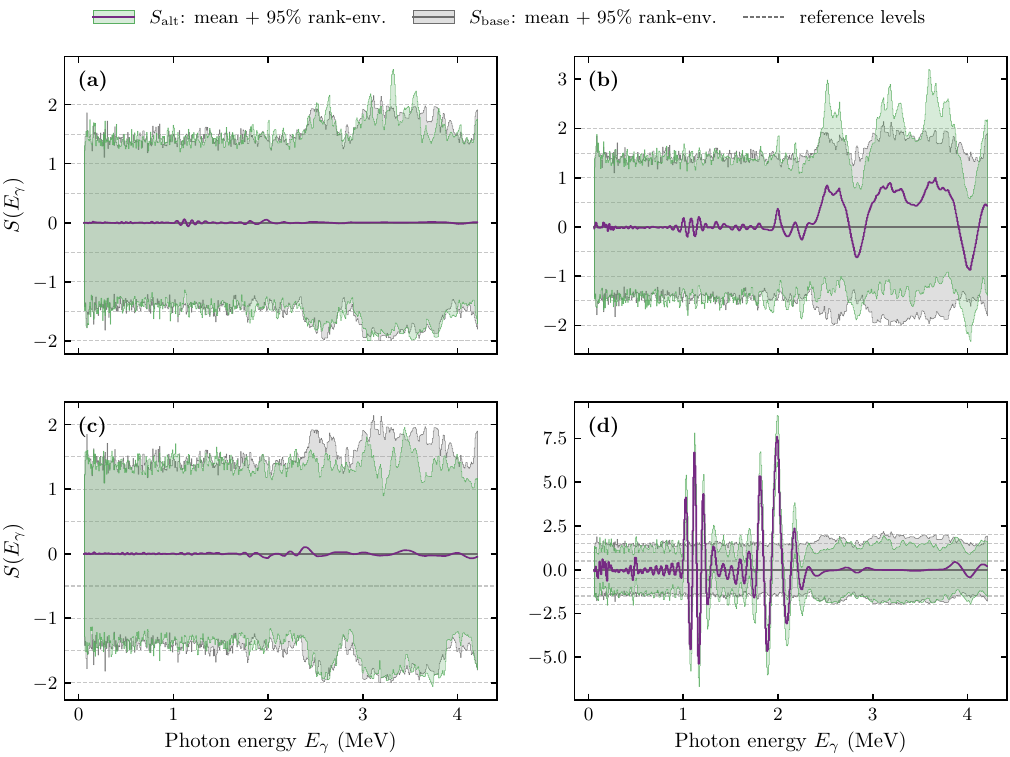}
  \caption{Robustness to alternative modeling choices for the
  $E_x\approx4.0$~MeV validation case. Panels show (a) a matched lognormal
  prior, (b) a constant $\bm{\mu}$-space prior center with the baseline
  RL-adaptive width schedule, (c) a fixed background expectation based on the
  Gamma-Poisson posterior mean, and (d) a response mismatch in which the synthetic data are generated with the OSCAR response normalized to a Gaussian standard deviation $\sigma_\gamma=30$~keV at $E_\gamma=1330$~keV, while the unfolding model uses $\sigma_\gamma=35$~keV at the same normalization energy. Each panel has its own vertical scale. The response-mismatch case produces the dominant deviation.}
  \label{fig:robustness}
\end{figure*}

We next examine how much the posterior changes when the empirical-Bayes prior or selected parts of the statistical model are modified. All comparisons in this subsection use the high-statistics $E_x\approx4.0$~MeV case. To quantify the change, we use the scaled-deviation metric defined in Sec.~\ref{subsec:sensitivity-robustness}. The empirical-Bayes sensitivity tests are separated into changes of the RL reference iteration (Fig.~\ref{fig:rl-sensitivity}) and changes of the adaptive prior-width schedule (Fig.~\ref{fig:sigma-sensitivity}), while the model-robustness tests are collected in Fig.~\ref{fig:robustness}. In each panel, the alternative posterior is shown through $S_{\mathrm{alt}}$, obtained by subtracting paired baseline posterior draws from alternative posterior draws and normalizing by the baseline 95\% rank-envelope half-width.

The baseline reference $S_{\mathrm{base}}$ is obtained by subtracting a random permutation of the baseline posterior draws from the original baseline draws. The gray band shows the scale of the baseline posterior variability under the same normalization, while the horizontal zero line marks no shift relative to the baseline posterior mean. The reference levels at $|S|=\{0.5,1.0,1.5,2.0\}$ illustrate displacements in units of the baseline half-width.

The mean curve of $S_{\mathrm{alt}}$ gives the shift of the alternative mean relative to the baseline posterior mean. A value of $S_{\mathrm{alt}}=\pm1.0$ corresponds to a displacement of one baseline half-width. The width and asymmetry of the colored band describe the distribution of the scaled differences between paired alternative and baseline posterior draws, while the gray band provides the corresponding baseline reference. By construction, the baseline difference distribution is symmetric around zero in the large-sample limit. In contrast, $S_{\mathrm{alt}}$ can be shifted, broadened, narrowed, or asymmetric depending on how the alternative posterior differs from the baseline.

The first test evaluates the posterior sensitivity to the number of iterations $t_{\mathrm{RL}}$ used to obtain the RL reference estimate. The automatic semi-convergence rule selected $t_{\mathrm{RL}}^{\mathrm{auto}}=58$ for this spectrum. For the under-iterated case, $t_{\mathrm{RL}}=5$ in Fig.~\ref{fig:rl-sensitivity}(a), the mean of $S_{\mathrm{alt}}$ remains within $|S|<1.0$, so its displacement from the baseline posterior mean is smaller than one baseline half-width. In the low-count region above $2.0$~MeV, the mean of $S_{\mathrm{alt}}$ exhibits some alternating variations around the zero-deviation line, while the scaled envelope shows minor changes in the upper and lower band edges relative to the baseline reference. This structure arises because the under-iterated prior gives an overly smooth intensity scale across multiple bins. While the local count information is insufficient to alter this prior shape strongly, the collective likelihood constraint still requires the integrated intensity over the high-$E_\gamma$ tail to match the observed count level. As a result, the posterior mean alternates above and below the baseline to some extent. This redistribution of intensity occurs for the low-count peaks at $E_\gamma\approx2.8$~MeV and $E_\gamma\approx4.0$~MeV, which are visible in Fig.~\ref{fig:prior-vs-post}. Intensity is reduced in those structures and shifted into the neighboring region with extremely few counts. However, the overall deviation of the alternative posterior remains smaller than the baseline posterior half-width.

For the over-iteration case, $t_{\mathrm{RL}}=500$ in Fig.~\ref{fig:rl-sensitivity}(b), the alternative mean curve remains within $|S|<0.5$ across the entire spectrum. The scaled envelope follows the baseline one, except for two smaller regions in the high-$E_\gamma$ tail, where the alternative upper band edge drops by more than 1.0 units. Inspecting the underlying prior structures reveals that these localized drops correspond exactly to bins where the over-iterated RL estimate has sudden drops in intensity, causing the upper edge of the prior band to shrink. Because this tail region is weakly constrained by the data, this prior configuration leads to some contraction of the posterior band.

Figure~\ref{fig:sigma-sensitivity} demonstrates how the posterior distribution responds to changes in the adaptive prior-width bounds, isolating the differing regularizing roles of $\sigma_{\min}$ and $\sigma_{\max}$ in the unfolding domain. In Fig.~\ref{fig:sigma-sensitivity}(a), applying a uniform $\sigma_j=1.0$ schedule for all bins, produces the most notable changes in the scaled deviation. In the signal-rich region, the alternative posterior mean shows only minor fluctuations below $|S|=0.5$. At higher $E_\gamma$, the mean of $S_{\mathrm{alt}}$ is larger, but it does not extend to the baseline half-width at $S=1.0$. This behavior occurs because restricting the width parameter to a uniform unit scale gives lower prior spread in the tail. In turn, this leads to a slight narrowing of the posterior band within this less data-informative region, where the prior influence is stronger. The effect is similar to that observed for the under-iterated RL case. Again, the alternative posterior has lower intensity within the low-statistics peaks at $E_\gamma\approx2.8$~MeV and $E_\gamma\approx4.0$~MeV in comparison with the baseline, while elevating the count level in the intermediate region.

In Fig.~\ref{fig:sigma-sensitivity}(b), increasing the maximum prior width to $\sigma_{\max}=5.0$, results in only minor fluctuations around the zero-deviation line, all bounded by the $|S|=0.5$ reference levels. Above $2.0$~MeV where the count levels are low, we observe that the lower edge of the $S_{\mathrm{alt}}$ envelope shifts downward relative to the baseline. The reason is that a larger lognormal $\sigma$ places more prior mass at very small fractional intensity levels, closer to the non-negativity boundary. This allows the lower edge of the alternative envelope to extend further downward, altering the asymmetry across this region. By permitting the extremely low-intensity region to drop close to the zero threshold, more intensity is effectively retained within the local spectral structures at $E_\gamma\approx2.8$~MeV and $E_\gamma\approx4.0$~MeV.

Modifying the minimum bound through a uniformly broader $\sigma_j=3.0$ schedule in Fig.~\ref{fig:sigma-sensitivity}(c) or a lower $\sigma_{\min}=0.3$ in Fig.~\ref{fig:sigma-sensitivity}(d) induces minimal variations, confirming that the unfolded spectrum is stable against these changes. For completeness, the sensitivity to the Gamma shape parameter $\alpha$ is reported in App.~\ref{app:alpha-sensitivity}, which yields negligible changes relative to the baseline posterior.

Next, we investigate the stability of the posterior under structural modifications to the underlying model. As shown in Fig.~\ref{fig:robustness}, replacing the baseline Gamma-lognormal hierarchy with a matched lognormal prior in panel (a) yields a negligible change in the mean of $S_{\mathrm{alt}}$ and a scaled envelope that remains close to the baseline. To isolate the effect of the prior distribution family, the alternative lognormal is constructed to preserve the mean and spread of the baseline prior. This choice represents a closely matched test between two similar heavy-tailed prior families. Transitioning from the baseline hierarchy to a fundamentally different prior family would represent a more profound structural modification. Nevertheless, the minimal deviation observed for this high-statistics spectrum indicates that the unfolded posterior is robust to replacing the baseline Gamma-lognormal hierarchy by this matched lognormal alternative.

In Fig.~\ref{fig:robustness}(b), using a constant prior center set to the average of the RL estimate can be viewed as a more extreme removal of local RL-shape information than the under-iterated case in Fig.~\ref{fig:rl-sensitivity}(a). Both the mean and envelope of $S_{\mathrm{alt}}$ follow a similar pattern, but the deviations are larger. The alternative posterior mean comes close to being one baseline half-width away from the baseline posterior mean for a few bins, and the corresponding shifts in the alternative posterior rank-envelope are larger than before.

Fixing the background expectation to the conjugate Gamma-Poisson posterior mean has a minimal impact on the mean of $S_{\mathrm{alt}}$ in Fig.~\ref{fig:robustness}(c). However, the upper edge of its envelope is reduced for most bins in the low-count region, with a reduction of up to one baseline half-width. Because background fluctuations constitute a key part of the total variance in the weak-signal regime, removing this source of uncertainty shrinks the permitted upward posterior variations.

Fig.~\ref{fig:robustness}(d) examines a response-operator mismatch. The synthetic data are generated with the OSCAR response normalized with Gaussian standard deviation $\sigma_\gamma=30$~keV at $E_\gamma=1330$~keV, while the unfolding model uses $\sigma_\gamma=35$~keV at the same normalization energy. This case produces the largest deviation among all tested configurations. Within the signal-rich region and around the highest peak structures, the mean scaled deviation undergoes large oscillations ranging from approximately $-5.0$ to $7.0$ baseline half-widths. Crucially, the rank-envelope of $S_{\mathrm{alt}}$ is too narrow to account for these deviations. The response mismatch is a systematic error for which the Poisson inverse problem itself changes, highlighting the importance of using a well-calibrated response operator.

Collectively, these high-statistics sensitivity and robustness studies show that the posterior is only modestly affected by the tested prior and background-model variations. Changing the Gamma shape parameter, the RL iteration, the prior-width bounds, the matched lognormal alternative, or the background treatment yields at most modest shifts on the scale of the baseline posterior uncertainty. In contrast, a mismatch in the $\gamma$-energy resolution operator used for synthetic data generation and unfolding changes the effective forward model and produces much larger deviations. A natural extension of the Bayesian framework is therefore to parameterize and include response-calibration uncertainties directly within the model, allowing this source of systematic uncertainty to be fully propagated into the posterior bands.

\subsection{Low-statistics validation}
\label{subsec:results-low-statistics}

We now evaluate the method under reduced count statistics. The low-statistics dataset is generated by reducing the total signal scale to 1\% of the high-statistics setting and increasing the background-to-signal ratio to $\rho=0.50$ from $\rho=0.15$. Since the background expectation is proportional to the expected detected signal, the absolute background counts are also much smaller than in the high-statistics dataset. The larger value of $\rho$ changes the relative composition of the ON counts, but the main change is that both the signal and background observations have much lower absolute counts. This setting tests the unfolding model in a genuinely low-count ON/OFF regime.

\begin{figure*}
  \centering
  \includegraphics{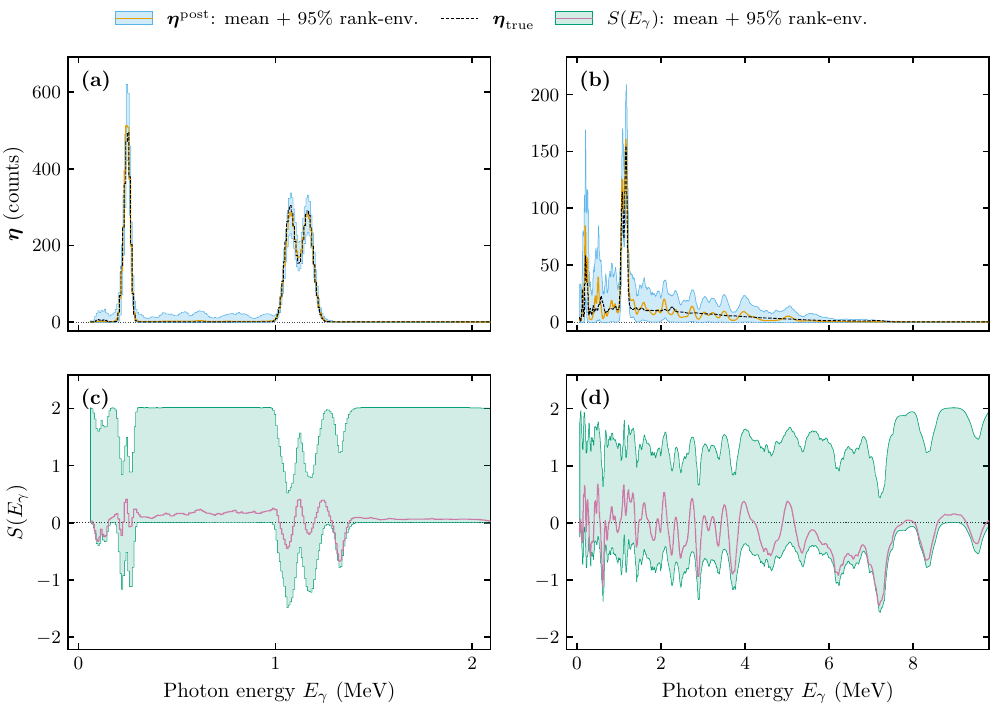}
  \caption{Low-statistics unfolded posterior in the resolution-limited space. The synthetic data use 1\% of the high-statistics signal scale and a background-to-signal ratio $\rho=0.50$. (a), (b) Posterior means and 95\% global rank-envelope bands for $E_x\approx2.5$~MeV and $E_x\approx9.5$~MeV, together with the ground truth $\bm{\eta}_{\mathrm{true}}$ as a black dashed curve. (c), (d) Corresponding scaled deviations relative to the truth, normalized by the posterior-envelope half-width.}
  \label{fig:multi-ex}
\end{figure*}

Panels (a) and (b) of Fig.~\ref{fig:multi-ex} show posterior results for $E_x\approx2.5$~MeV and $E_x\approx9.5$~MeV spectra. In both cases, the posterior bands give a reasonable uncertainty description relative to the known ground truth. The increase in uncertainty compared with the high-statistics case is a consequence of the reduced absolute count levels. At $E_x\approx2.5$~MeV, the resolution-limited truth is dominated by a small number of pronounced peak structures. These should not be interpreted as individual emitted lines. As seen in Fig.~\ref{fig:synthetic-data}, the underlying $\bm{\mu}_{\mathrm{true}}$ spectrum contains several $\delta$-like peaks that are broadened and partly merged by the detector-resolution mapping before appearing in $\bm{\eta}$-space. The posterior mean follows these resolution-limited peak structures and returns to a low non-negative intensity in the intermediate regions. 

At $E_x\approx9.5$~MeV, the lower-$E_\gamma$ peak structures are accompanied by a weak positive component that extends into the high-$E_\gamma$ tail. This component is the resolution-limited contribution from numerous smaller emitted transitions in the quasi-continuum, not a single smooth background-like feature. On the absolute scale, the posterior band in this tail appears narrow because the underlying intensity is small. 

The scaled-deviation panels (c) and (d) of Fig.~\ref{fig:multi-ex} reveal the same non-negativity boundary behavior as discussed for the high-statistics case in Sec.~\ref{subsec:baseline-prior-4MeV}. In weak-signal regions, the posterior mean lies close to the lower edge of the rank-envelope, while the upper portion of the envelope represents the remaining upward posterior variation. For $E_x\approx2.5$~MeV, this occurs mainly where there is no true intensity. In the high-$E_\gamma$ tail of the $E_x\approx9.5$~MeV spectrum, where the truth is very small but non-zero, the posterior mean frequently falls below the true value. Under weak data constraints, the combined effect of the Poisson likelihood and the Gamma-lognormal prior hierarchy concentrates probability mass near the zero boundary, producing a highly skewed distribution with a long upward uncertainty tail. The resulting asymmetric 95\% rank-envelope reflects the one-sided uncertainty of a weakly constrained Poisson inverse problem.

\subsection{\boldmath Prior dependence in the low-statistics regime}
\label{subsec:results-lowstat-prior-dependence}
Since specific prior configurations are expected to influence the posterior more when the likelihood is weak, we perform an additional targeted prior-dependence check for the $E_x\approx9.5$~MeV spectrum. This spectrum is a useful low-statistics stress test because the lower-$E_\gamma$ structures coexist with a weak positive high-$E_\gamma$ tail. In this tail, the absolute resolution-limited intensity is very small and the posterior is close to the non-negativity boundary.

For this comparison, we do not use the scaled-deviation diagnostic as the main visualization. In Eq.~\eqref{eq:scaled-deviation-alt}, the difference between an alternative and the baseline posterior is divided by the baseline envelope half-width $w_{j,1/2}$. In the weak high-$E_\gamma$ tail of this low-statistics case, this half-width is itself very small in absolute count units. As a result, the posterior shifts of order one count or less can appear large after normalization by $w_{j,1/2}$. We therefore compare the baseline and alternative posteriors directly on the absolute count scale in $\bm{\eta}$-space, using a symmetric-logarithmic vertical axis.

\begin{figure*}
  \centering
  \includegraphics{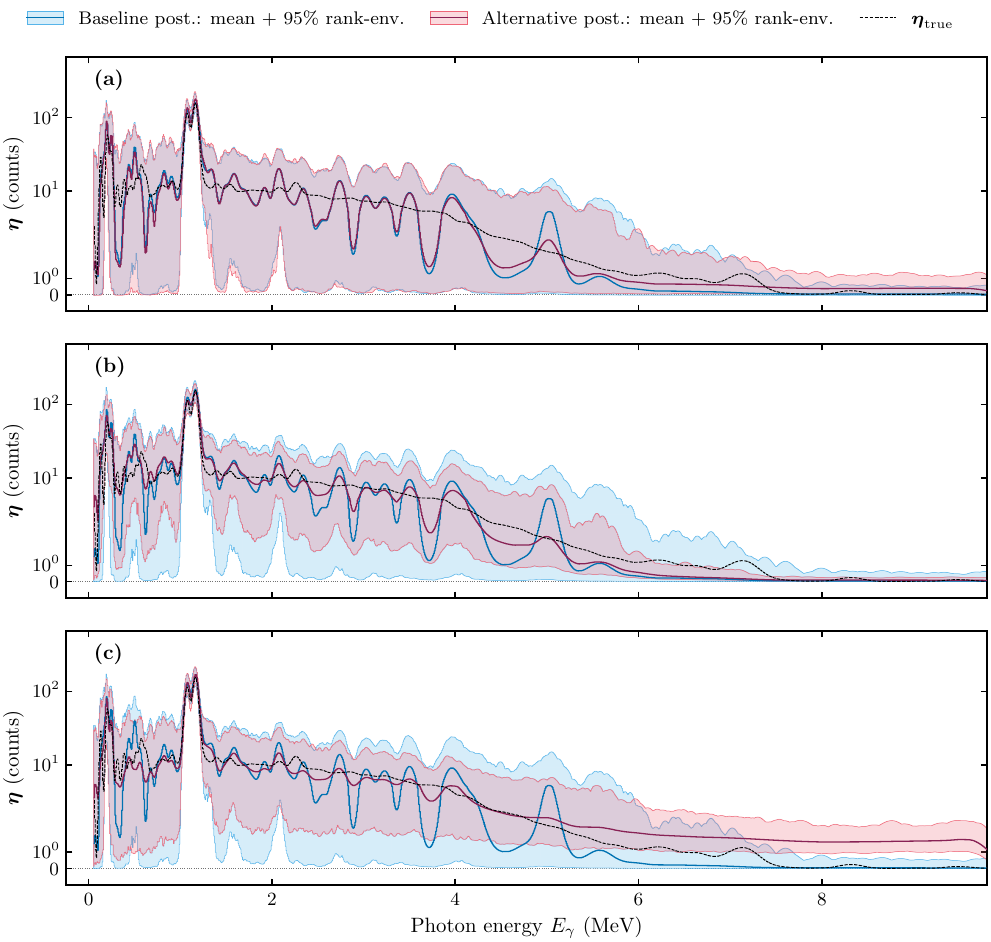}
  \caption{Targeted prior-dependence check for the low-statistics $E_x\approx9.5$~MeV spectrum, shown on the absolute resolution-limited count scale with a symmetric-logarithmic vertical axis. Each panel compares the baseline posterior with an alternative prior construction. The baseline is centered on the RL reference and uses the adaptive prior-width schedule with $(\sigma_{\min},\sigma_{\max})=(1.0, 3.0)$. (a) Constant prior center set to the mean of the RL estimate with the baseline adaptive $\sigma_j$ schedule. (b) Baseline prior center kept, but the adaptive schedule is replaced with a uniform $\sigma_j=1.0$ schedule. (c) Both changes combined as a stress test. Bands denote 95\% global rank envelopes, solid curves denote posterior means, and the black dashed curve displays the known synthetic truth $\bm{\eta}_{\mathrm{true}}$. The comparison is shown on the absolute count scale because, in the weak high-$E_\gamma$ tail, the baseline posterior half-width can be small in count units, so scaled deviations may magnify posterior shifts that are minor on the physical scale.}
  \label{fig:lowstat-prior-dependence}
\end{figure*}

Figure~\ref{fig:lowstat-prior-dependence} shows that the posterior is largely stable on the absolute count scale while also revealing where the empirical-Bayes construction affects the weak tail. In Fig.~\ref{fig:lowstat-prior-dependence}(a), replacing the local RL prior center by a constant center leaves the posterior mean nearly unchanged over the main part of the spectrum, up to about $E_\gamma\simeq4$--$5$~MeV. The visible differences occur mainly in the far high-$E_\gamma$ tail. Above roughly $E_\gamma\simeq6$~MeV, the alternative mean is slightly elevated relative to the baseline, but both posteriors remain at very small absolute count levels. Thus, although the displacement would be amplified in a scaled-deviation plot, the absolute change is very small compared with the scale of the dominant spectral structures.

Figure~\ref{fig:lowstat-prior-dependence}(b) isolates the effect of the adaptive width schedule by replacing it with a uniform $\sigma_j=1.0$ schedule. The alternative posterior mean also shows smaller oscillatory variations. This smoother behavior is more consistent with the slowly varying component of the truth spectrum. The oscillatory freedom allowed by the broader baseline schedule is not always needed locally, even though it is useful for avoiding over-constraint in weakly informed regions.

Figure~\ref{fig:lowstat-prior-dependence}(c) combines the constant prior center with the uniform $\sigma_j=1.0$ schedule. This is the most restrictive of the three alternatives and should be interpreted as a stress test. Even in this case, the posterior mean and band remain consistent with the truth through the main signal-bearing region, approximately up to $E_\gamma\simeq6$--$7$~MeV. At higher $E_\gamma$, the lower edge of the alternative posterior band is lifted to a small positive level while the truth is close to zero. This produces a noticeable difference relative to the baseline when measured in units of the baseline half-width, but on the absolute count scale the difference is again small.

Summarizing, Fig.~\ref{fig:lowstat-prior-dependence} should then be interpreted as showing a flexibility-smoothing trade-off in the low-statistics regime. It is not a simple ranking of the prior choices. In this $E_x\approx9.5$~MeV spectrum, the alternatives with a uniform $\sigma_j=1.0$ schedule attenuate oscillatory variations of the baseline posterior mean over an extended part of the spectrum, where the truth varies slowly. In these regions, the smoother alternative posteriors can follow the truth more closely than the more flexible baseline posterior. The difference changes character in the extreme high-$E_\gamma$ tail. There, the truth is at the sub-count level or close to zero, and the combined constant-center and uniform-width stress test lifts the posterior band to a small positive floor. This produces a visible difference relative to the baseline when measured in units of the baseline half-width, but the absolute difference remains small on the count scale. In this way, the baseline prior is best viewed as a practical default compromise across spectra. It leaves more local freedom in weakly informed regions, which can allow some oscillatory variation in slowly changing parts of this particular spectrum, but it also reduces the risk of over-constraining near-boundary tails where the likelihood carries little information.

\subsection{Comparison with the frequentist unfolding method}
\label{subsec:results-comparison}

\begin{figure*}
  \centering
  \includegraphics[width=0.85\textwidth]{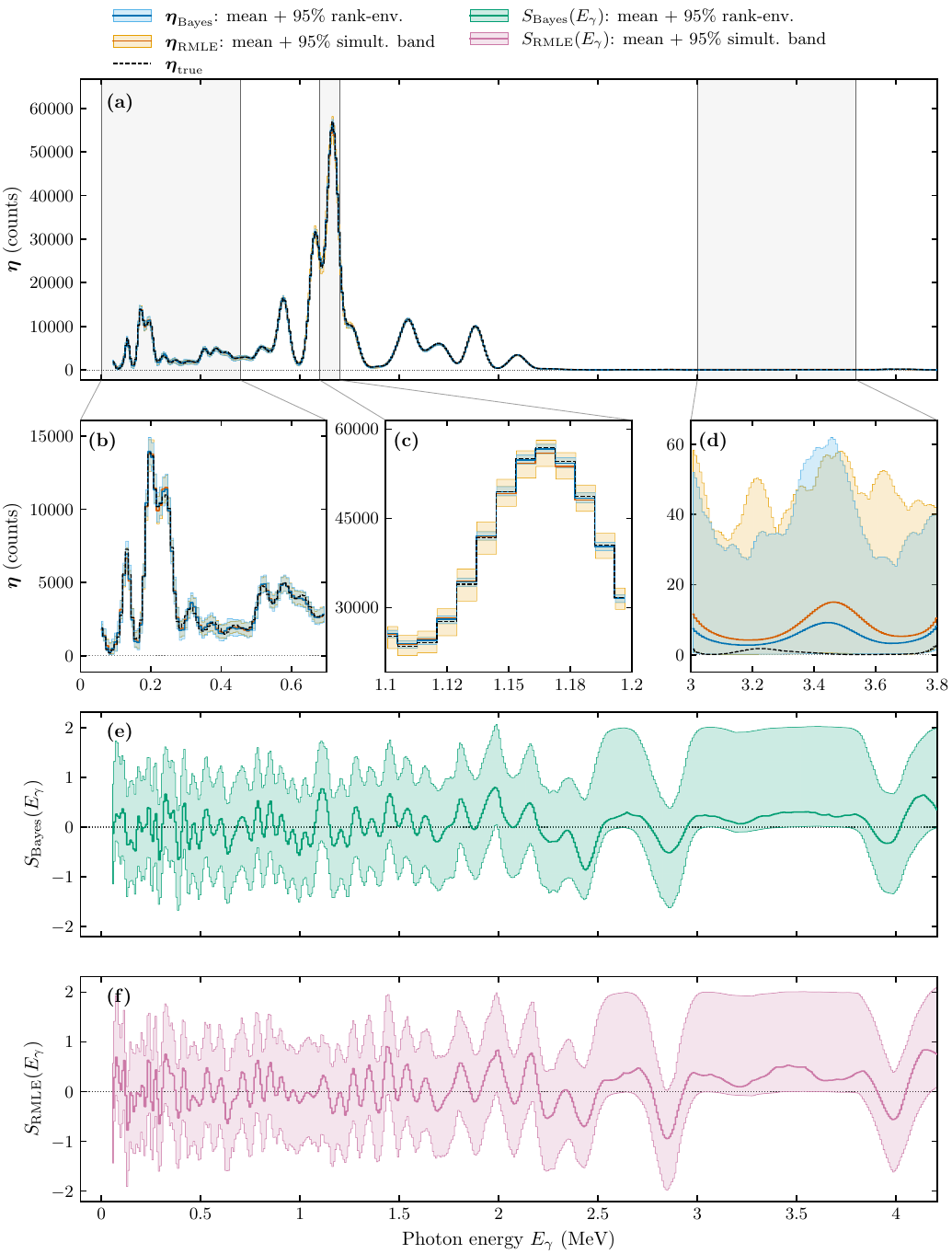}
  \caption{Bayesian-frequentist comparison for the high-statistics $E_x\approx 4.0$~MeV case. (a) Full unfolded spectrum in $\bm{\eta}$-space. (b)--(d) Local zooms of representative spectral structures. (e), (f) Method-specific scaled deviations relative to the synthetic truth, where zero corresponds to $\bm{\eta}_{\mathrm{true}}$. The Bayesian result is summarized by the posterior mean and a 95\% global rank-envelope. The RMLE result uses the sparsity-penalized fit with the penalty strength selected by minimizing the Wasserstein-1 distance to the known synthetic truth in $\bm{\eta}$-space, and is summarized by the bootstrap mean and a simultaneous 95\% Bonferroni-percentile bootstrap band~\cite{lima2025}.}
  \label{fig:bayes-vs-freq-high}
\end{figure*}

\begin{figure*}
  \centering
  \includegraphics[width=0.85\textwidth]{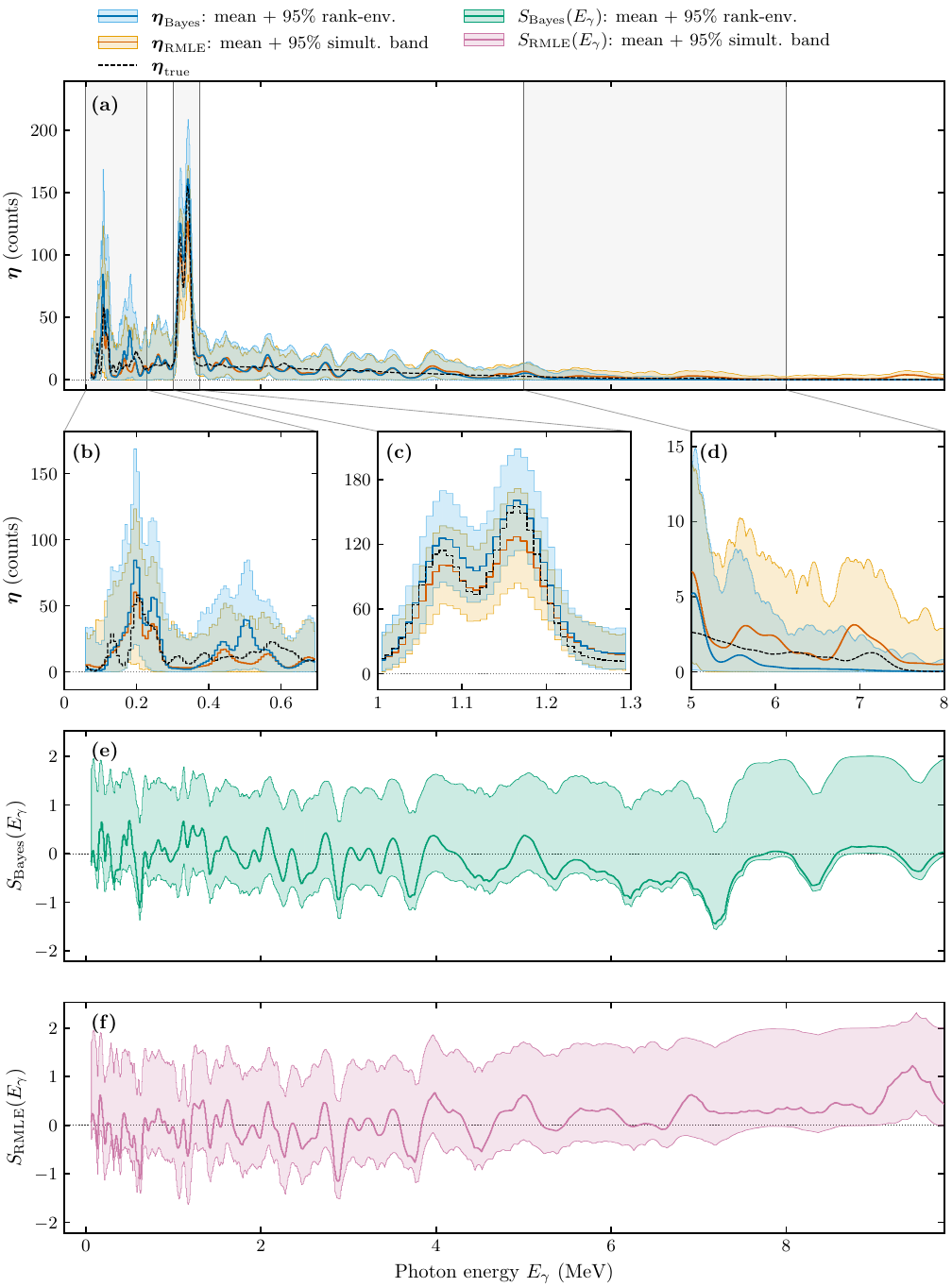}
  \caption{Bayesian-frequentist comparison for the low-statistics $E_x\approx 9.5$~MeV case. The layout follows Fig.~\ref{fig:bayes-vs-freq-high}: (a) full unfolded spectrum in $\bm{\eta}$-space, (b)--(d) local zooms, and (e), (f) method-specific scaled deviations relative to $\bm{\eta}_{\mathrm{true}}$. The zoom panels highlight the low-energy structure, the merged double-peak region, and the high-$E_\gamma$ tail. The RMLE result in this case uses the unpenalized fit and is summarized by the bootstrap mean and a simultaneous 95\% Bonferroni-percentile bootstrap band.}
  \label{fig:bayes-vs-freq-low}
\end{figure*}

Finally, we compare the Bayesian posterior results with the frequentist RMLE framework recently developed for unfolding $\gamma$-ray spectra in nuclear-physics applications~\cite{lima2025}. The two spectra used here are not new validation cases. The $E_x\approx 4.0$~MeV spectrum is the high-statistics case analyzed above, and the $E_x\approx9.5$~MeV spectrum is the low-statistics case from Fig.~\ref{fig:multi-ex}. We therefore do not repeat the Bayesian validation here, but focus on how the Bayesian and RMLE uncertainty summaries compare for the same synthetic observations.

For the high-statistics comparison, the RMLE fit uses the OMpy sparsity penalty, with the penalty strength selected by minimizing the Wasserstein-1 distance to the known synthetic truth in $\bm{\eta}$-space. This truth-based selection is possible only because the comparison is performed on synthetic data, and it should be interpreted as a controlled benchmark choice, not as a prescription for experimental data. For the low-statistics comparison, we use the unpenalized RMLE fit. In both cases, the RMLE uncertainty band is obtained from 1000 bootstrap resamples and summarized by a simultaneous 95\% Bonferroni-percentile band.

The Bayesian and frequentist uncertainty bands have different statistical meanings. The Bayesian global rank-envelope summarizes posterior uncertainty about $\bm{\eta}$ for the fixed observed dataset, conditional on the assumed model and empirical-Bayes prior construction. The RMLE bootstrap band summarizes the sampling variability of the RMLE estimator under repeated resampling. The scaled-deviation panels are therefore normalized separately for each method, using that method's own half-width, and the two band widths should not be understood as estimates of the same uncertainty object.

Figure~\ref{fig:bayes-vs-freq-high} displays the high-statistics comparison at $E_x\approx4.0$~MeV. The Bayesian posterior mean and the sparsity-penalized RMLE bootstrap mean are very similar over the full spectrum and in the local zoom panels. The clearest visible difference is in the largest peak region, especially in Fig.~\ref{fig:bayes-vs-freq-high}(c), where the RMLE bootstrap band is broader than the Bayesian band. The method-specific scaled-deviation panels show nearly the same structure for both methods.

The low-statistics comparison at $E_x\approx9.5$~MeV is given in Fig.~\ref{fig:bayes-vs-freq-low}. The deviations of the Bayesian posterior mean and the RMLE bootstrap mean from the truth alternate across the spectrum rather than favoring one method systematically. In the merged double-peak region in Fig.~\ref{fig:bayes-vs-freq-low}(c), the RMLE mean lies slightly closer to the truth on parts of the first peak and the intervening dip, while the Bayesian posterior mean is closer on parts of the rise toward the second peak. In the high-$E_\gamma$ tail, the RMLE band is broader, while both bands remain compatible with the truth on the scale shown.

The comparison gives the same qualitative picture in both count regimes. The Bayesian and RMLE procedures give similar resolution-limited unfolded spectra, while attaching different statistical meanings to the uncertainty bands. The Bayesian framework should therefore be viewed as a complementary posterior formulation of Oslo-method unfolding, not as a replacement for the frequentist RMLE approach.

\section{Conclusion}
\label{sec:conclusion}

In this work, we developed and validated an empirical-Bayes hierarchical framework for unfolding $\gamma$-ray spectra in the Oslo method. By formulating the problem as posterior inference under a Poisson forward model, the method provides non-negative unfolded spectra in the resolution-limited space together with simultaneous uncertainty bands and predictive diagnostics. The framework addresses several challenges in Oslo-type unfolding, including uncertainty quantification, stable treatment of low-count regions, and probabilistic incorporation of background through a joint ON/OFF model.

Validation on synthetic data shows that the method recovers the resolution-limited spectrum within the reported uncertainty bands in both high- and low-statistics cases considered here. The hierarchical prior remains broad in weakly constrained regions while allowing substantial posterior contraction in signal-rich regions. In the high-statistics case, the posterior is stable under the tested changes to the empirical-Bayes construction and under the matched-prior family and background-treatment alternatives. These tests include changes to the Richardson-Lucy iteration, the prior-width bounds, the Gamma shape parameter, and the prior center. The additional low-statistics prior-dependence check shows a more nuanced behavior: changing either the prior-width schedule or the RL-based prior center alone produces only moderate shifts, whereas changing both simultaneously produces the largest scaled deviations in the weak high-$E_\gamma$ tail. On the absolute scale, these differences remain small compared with the dominant spectral structures. The robustness study also identifies the detector-response model as the most consequential modeling choice among those tested here. A mismatch in the assumed $\gamma$-energy resolution produces large structured deviations in the unfolded spectrum.

A direct comparison with the recent frequentist RMLE formulation of Ref.~\cite{lima2025} shows that the two approaches yield consistent unfolded spectra for the representative high- and low-statistics cases considered here. We therefore do not conclude that the Bayesian formulation systematically outperforms the frequentist one in terms of unfolded accuracy. Rather, the two methods provide comparable physical reconstructions with different inferential meanings. The Bayesian bands describe posterior uncertainty about the unknown spectrum for the fixed observed dataset and assumed model, while the RMLE bootstrap bands describe the repeated-sampling variability of the estimator. In this sense, the Bayesian method is best viewed not as a replacement for RMLE, but as a complementary formulation of the same inverse problem. Its main strength is the unified posterior treatment of regularization, background, predictive checking, and uncertainty propagation. In particular, it establishes a route toward a more probabilistic Oslo method in which uncertainties from the unfolding step can be carried forward to the extraction of first-generation spectra, nuclear level densities, and $\gamma$-ray strength functions. The frequentist method, on the other hand, is computationally lighter in the implementations considered here, and is more efficient for matrix-scale unfolding.

\section{Outlook}
\label{sec:outlook}
A natural next step is to include the detector-response uncertainty directly in the Bayesian model. The robustness study shows that a mismatch in the assumed $\gamma$-energy resolution can produce substantially larger deviations than the empirical-Bayes choices tested here. In the present work, the response operator is fixed during inference. Future work should therefore explore parameterizations in which response-width or response-shape parameters are inferred jointly with the unfolded spectrum, or marginalized over using calibration constraints. This would allow response-calibration uncertainty to be propagated into the final unfolded bands rather than being treated as a separate systematic check.

A second methodological improvement concerns the role of the $\gamma$-energy discretization in the unfolding process. The present implementation uses a fixed $E_\gamma$ grid, which is convenient and compatible with the existing Oslo-method workflow. However, because the detector resolution changes with $E_\gamma$, a fixed fine bin width can overparameterize parts of the spectrum where the instrumental resolution is comparatively broad. This can produce highly correlated response columns and strong posterior correlations, especially in weakly constrained high-$E_\gamma$ regions.

Future work should explore representations better matched to the detector's energy-dependent resolution, such as adaptive binning or multi-resolution parameterizations. Ideally, the number of bins per local full width at half maximum should remain consistent, avoiding over-resolution in regions where the detector cannot resolve such fine structure. Implementing such a scheme within the Oslo framework is non-trivial, because any non-uniform representation must eventually be mapped back to the uniform grid used by subsequent steps of the method without introducing artificial smoothing. Statistically controlled mapping operators that preserve total counts and account for induced correlations would benefit both the Bayesian framework and the frequentist RMLE formulation~\cite{lima2025}. Future progress in Oslo-method unfolding is likely to depend not only on the choice of Bayesian or frequentist inference, but also on improving the statistical representation of the spectrum itself.

\acknowledgments

Simulations with RAINIER were performed on resources provided by Sigma2, the National Infrastructure for High Performance Computing and Data Storage in Norway (using ``Saga'' on Project No. NN9464K). 
A.~C.~L.\ and E.~L.\ gratefully acknowledge funding of this research by the Research Council of Norway, Project Grant No. 316116. A.~C.~L.,  E.~L.\ and A.~H.~M.\ acknowledge support from the Research Council of Norway through the Norwegian Nuclear Research Centre (project No. 341985). A.~K.\ was supported by the Research Council of Norway through the FRIPRO grant 323985 PLUMBIN’.
All authors also acknowledge continued support through dScience (Centre for Computational and Data Science) at the University of Oslo, Norway.
We are grateful to Lasse Lorentz Braseth, Dr.\ Fabio Zeiser and Dr.\ Maria Markova for stimulating discussions and inspiring comments.

\section*{Data availability}

The source code, configuration files, input data, and generated result files required to reproduce the synthetic-data generation, Bayesian unfolding results, frequentist RMLE comparisons, benchmark summaries, and all figures in this article are available in the Zenodo archive in Ref.~\cite{mjos2026archive}. The corresponding development version of the code is available at \url{github.com/andreashmj/empirical-bayes-unfolding-oslo-method}.

\appendix
\section{Sampler backend benchmarks}
\label{app:benchmarks}

The Bayesian unfolding posterior was sampled using NUTS backends compatible with the \texttt{PyMC} model specification. The timing comparison reported below includes the standard \texttt{PyMC} NUTS implementation~\cite{pymc2023}, \texttt{NumPyro}~\cite{phan2019numpyro}, and \texttt{BlackJAX}~\cite{cabezas2024blackjax}. The purpose of the benchmark is to compare practical wall-clock performance for the same statistical model, likelihood, prior construction, response matrices, and synthetic datasets. The benchmark was performed for three representative excitation-energy rows in both the high- and low-statistics synthetic settings.

All reported runs use double-precision arithmetic. This was necessary for stable NUTS sampling in the present high-dimensional unfolding problem. GPU execution was therefore not included in the final comparison. The available GPU hardware was not optimized for double-precision arithmetic, and the tested GPU configuration was slower than the CPU runs once the model was forced to run in \texttt{float64}. The CPU results in Table~\ref{tab:benchmarks} should therefore be interpreted as hardware-dependent timing comparisons. 

A sampler backend was included in the timing comparison only if it satisfied the diagnostic stability criterion for all benchmark runs under the fixed configuration used here. Specifically, the repeated runs were required to show no systematic divergent transitions, rank-normalized split-$\widehat{R}$ values below 1.01, and high bulk ESS. The \texttt{Nutpie} backend~\cite{seyboldt2026nutpie} was also tested, but it did not satisfy the full stability criterion and is therefore omitted from Table~\ref{tab:benchmarks}. This should be interpreted only as a result for this model, initialization, target-acceptance schedule, and backend configuration, not as a general assessment of the sampler backend. The high-statistics $E_x\approx9.5$~MeV \texttt{PyMC} row contains one divergent transition in the maximum-over-repetitions diagnostic, but this isolated event was not accompanied by elevated $\widehat{R}$ or low ESS.

Among the reported benchmark runs, no single backend is uniformly fastest. The JAX-based backends reduce wall-clock time relative to native \texttt{PyMC} in several cases, but their relative performance depends on the active spectrum size and count regime.

\begin{table*}[t]
\caption{Sampling performance comparison for selected excitation-energy bins. All reported results use 4 parallel chains on 4 CPU cores, with 2000 tuning steps and 2000 posterior draws per chain. The entries are aggregated over three independent repetitions with different sampler seeds. Time is reported as the mean wall-clock duration and sample standard deviation over these repetitions, rounded to the nearest second. The standard deviation is a descriptive measure of run-to-run variability, not a precise uncertainty estimate. Div., MTD~(\%), and $\widehat{R}_{\max}$ report the maximum value over repetitions, while $\mathrm{ESS}_{\min}$ and $\mathrm{ESS}_{\min}/s$ are averaged over repetitions. Here Div.\ denotes the number of divergent transitions after tuning, MTD~(\%) is the percentage of post-warmup draws that reached the maximum tree depth, and $\widehat{R}_{\max}$ is the maximum rank-normalized split $\widehat{R}$ across all $E_\gamma$ bins. All tests were performed on an Intel Core i7-14700HX system with 20 cores and 28 hardware threads. The samplers were initialized near the Richardson-Lucy estimate with chain-specific jitter. Only sampler backends satisfying the diagnostic stability criterion across all benchmark runs are included.}
\label{tab:benchmarks}
\begin{ruledtabular}
\begin{tabular}{lcclccccccl}
Stats. & $E_x$ (keV) & $N_{\mathrm{bins}}$ & Sampler &
Target acc. & Div. & MTD (\%) & $\widehat{R}_{\max}$ &
$\mathrm{ESS}_{\min}$ & $\mathrm{ESS}_{\min}/s$ & Time (s) \\
\hline
High & 2500 & 255  & PyMC     & 0.99 & 0 & 0.0 & 1.003 & 2821 & 28.2 & $100 \pm 2$ \\
High & 2500 & 255  & NumPyro  & 0.99 & 0 & 0.0 & 1.001 & 2880 & 45.3 & $64 \pm 1$ \\
High & 2500 & 255  & BlackJAX & 0.99 & 0 & 0.0 & 1.004 & 2370 & 41.7 & $57 \pm 2$ \\
\hline
High & 4000 & 519  & PyMC     & 0.95 & 0 & 0.0 & 1.003 & 2646 & 2.9 & $917 \pm 13$ \\
High & 4000 & 519  & NumPyro  & 0.95 & 0 & 0.0 & 1.003 & 2639 & 3.8 & $698 \pm 51$ \\
High & 4000 & 519  & BlackJAX & 0.95 & 0 & 0.0 & 1.002 & 2485 & 4.0 & $620 \pm 35$ \\
\hline
High & 9500 & 1218 & PyMC     & 0.90 & 1 & 0.0 & 1.003 & 3274 & 1.9 & $1769 \pm 37$ \\
High & 9500 & 1218 & NumPyro  & 0.90 & 0 & 0.0 & 1.002 & 3603 & 2.1 & $1750 \pm 23$ \\
High & 9500 & 1218 & BlackJAX & 0.90 & 0 & 0.0 & 1.002 & 3577 & 1.8 & $1973 \pm 24$ \\
\hline
Low  & 2500 & 255  & PyMC     & 0.99 & 0 & 0.0 & 1.006 & 2402 & 19.1 & $127 \pm 23$ \\
Low  & 2500 & 255  & NumPyro  & 0.99 & 0 & 0.0 & 1.002 & 4703 & 58.2 & $81 \pm 2$ \\
Low  & 2500 & 255  & BlackJAX & 0.99 & 0 & 0.0 & 1.002 & 4528 & 64.2 & $70 \pm 2$ \\
\hline
Low  & 4000 & 519  & PyMC     & 0.95 & 0 & 0.0 & 1.003 & 2877 & 12.1 & $239 \pm 4$ \\
Low  & 4000 & 519  & NumPyro  & 0.95 & 0 & 0.0 & 1.002 & 3081 & 18.8 & $164 \pm 1$ \\
Low  & 4000 & 519  & BlackJAX & 0.95 & 0 & 0.0 & 1.002 & 3047 & 20.6 & $148 \pm 2$ \\
\hline
Low  & 9500 & 1218 & PyMC     & 0.90 & 0 & 0.0 & 1.002 & 2797 & 5.6 & $497 \pm 5$ \\
Low  & 9500 & 1218 & NumPyro  & 0.90 & 0 & 0.0 & 1.004 & 2531 & 4.9 & $535 \pm 99$ \\
Low  & 9500 & 1218 & BlackJAX & 0.90 & 0 & 0.0 & 1.003 & 2438 & 4.4 & $559 \pm 2$ \\
\end{tabular}
\end{ruledtabular}
\end{table*}
\clearpage

\section{Log-concavity and sampler behavior}
\label{app:concavity}
This appendix section summarizes the local concavity properties of the likelihood and prior that are relevant for the NUTS sampler behavior discussed in Sec.~\ref{subsec:nuts}. The results do not imply that the full hierarchical posterior is globally log-concave. However, they show that important conditional blocks of the model have favorable curvature properties for gradient-based sampling.

Ignoring terms independent of $\bm{\mu}$, the Poisson log-likelihood is
\begin{equation*}
\ell(\bm{\mu})
=\sum_i
\Big\{
n_i \log \big[(\mathbf{R}_\gamma \bm{\mu})_i \big] -
             (\mathbf{R}_\gamma \bm{\mu})_i
\Big\}.
\end{equation*}
If $\bm{a}_i^\tr$ denotes the $i$-th row of $\mathbf{R}_\gamma$, then
\begin{equation*}
(\mathbf{R}_\gamma \bm{\mu})_i=\bm{a}_i^\tr \bm{\mu},
\end{equation*}
and the Hessian can be written as 
\begin{equation*}
\nabla^2 \ell(\bm{\mu})= - \sum_i \frac{n_i}{(\bm{a}_i^\tr \bm{\mu})^2}\, \bm{a}_i \bm{a}_i^\tr
\leq \bm{0}.
\end{equation*}
Equivalently,
\begin{equation*}
\nabla^2 \ell(\bm{\mu})= - \mathbf{R}_\gamma^\tr
\operatorname{diag}\left(\frac{n_1}{\nu_1^2},\dots,\frac{n_J}{\nu_J^2}\right)
\mathbf{R}_\gamma\le \bm{0}.
\end{equation*}
Hence, the Poisson log-likelihood is concave on the positive domain. Adding a fixed non-negative background contribution preserves this concavity, since the logarithm is then applied to a positive function of $\bm{\mu}$. From Eq.~\eqref{eq:mu-prior}, the conditional prior on each emitted-spectrum bin is 
\begin{equation*}
\mu_j \mid m_j \sim \mathrm{Gamma}\left(\alpha,\beta_j=\frac{\alpha}{m_j}\right).
\end{equation*}
Up to terms independent of $\mu_j$, the corresponding conditional log-density is 
\begin{equation*}
\log \pi(\mu_j\mid m_j) = (\alpha-1)\log \mu_j -
\alpha \mu_j/m_j +
\text{const}(m_j,\alpha),
\end{equation*}
with second derivative
\begin{equation*}
\partial_{\mu_j}^2 \log \pi(\mu_j\mid m_j) =
- \frac{\alpha-1}{\mu_j^2}.
\end{equation*}
Thus, for $\alpha\ge 1$, the conditional prior is log-concave in $\mu_j$, and for $\alpha>1$ it is strictly log-concave. It follows that, conditional on the latent mean layer $\mathbf{m}$, the posterior
\begin{equation*}
\log \pi(\bm{\mu}\mid \mathbf{m},\mathbf{n})=\ell(\bm{\mu})+\log \pi(\bm{\mu}\mid \mathbf{m}),
\end{equation*}
is concave in $\bm{\mu}$ when $\alpha\ge 1$. For the latent mean layer, define
\begin{equation*}
z_j=\log m_j.
\end{equation*}
Equation~\eqref{eq:mu-prior} gives
\begin{equation*}
z_j \sim \Normal(\zeta_j,\sigma_j^2),
\qquad
\zeta_j=\log \mu_{\mathrm{RL},j}-\frac{1}{2}\sigma_j^2.
\end{equation*}
Conditioning on $\mu_j$, the log-density in $z_j$ is 
\begin{equation*}
\log \pi(z_j\mid \mu_j) =
-\alpha z_j
-\alpha \mu_j e^{-z_j}
-\frac{(z_j-\zeta_j)^2}{2\sigma_j^2}
+ \text{const},  
\end{equation*}
and therefore
\begin{equation*}
\partial_{z_j}^2 \log \pi(z_j\mid \mu_j) =
-\alpha \mu_j e^{-z_j}
-\frac{1}{\sigma_j^2} <0.
\end{equation*}
Accordingly, the mean layer is strictly log-concave in $z_j$. For $\alpha\ge 1$ the model is blockwise log-concave. The posterior is concave in $\bm{\mu}$ conditional on $\mathbf{m}$, and strictly log-concave in $\bm{z}=\log \mathbf{m}$ conditional on $\bm{\mu}$. The joint posterior need not be globally concave because the Gamma layer couples $\mu_j$ and $m_j$ through the nonlinear term  $-\alpha \mu_j/m_j$. Nevertheless, this blockwise structure gives favorable conditional curvature and can give stable gradients in practice, which is useful for HMC/NUTS. When $\alpha>1$, the term $(\alpha-1) \log\mu_j$ also induces a soft barrier as $\mu_j\to 0^+$. For $\alpha=1$, this extra barrier disappears, although the support remains restricted to $\mu_j>0$. For $0<\alpha<1$, the log-concavity argument for the $\mu_j$-block no longer applies.

\section{Sensitivity to the prior shape parameter}
\label{app:alpha-sensitivity}

\begin{figure}
  \centering
  \includegraphics{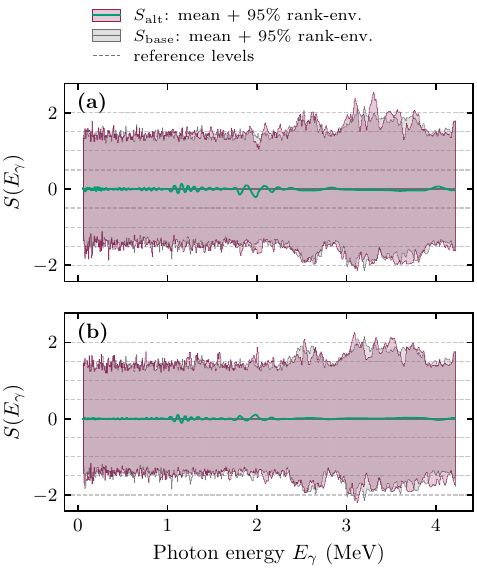}
  \caption{Posterior sensitivity to the Gamma shape parameter $\alpha$, quantified by the scaled deviation $S_{\mathrm{alt}}$. Panels show (a) $\alpha=0.3$ and (b) $\alpha=3.0$, compared with the baseline $\alpha=1.0$. The close overlap between the alternative and the baseline null distribution indicates that the posterior is practically insensitive to these changes in $\alpha$.}
  \label{fig:alpha-sensitivity}
\end{figure}

We examine the sensitivity to the Gamma shape parameter $\alpha$ for the high-statistics $E_x\approx 4.0$~MeV case by comparing the baseline $\alpha=1.0$ with two extreme alternatives, $\alpha=0.3$ and $\alpha=3.0$. The resulting scaled deviations remain close to the baseline null distribution over the full $E_\gamma$ range. We therefore find no practical sensitivity to $\alpha$ in this case.

% - Bibliography (BibTeX) -
\bibliographystyle{apsrev4-2}
\bibliography{bibliography}

@incollection{neal2011mcmc,
  author    = {Neal, Radford M.},
  title     = {{MCMC} Using Hamiltonian Dynamics},
  booktitle = {Handbook of Markov Chain Monte Carlo},
  editor    = {Brooks, Steve and Gelman, Andrew and Jones, Galin L. and Meng, Xiao-Li},
  year      = {2011},
  publisher = {Chapman \& Hall/CRC},
  address   = {Boca Raton, FL},
  pages     = {113--162},
  doi       = {10.1201/b10905-6}
}

@article{hoffman2014nuts,
  author  = {Hoffman, Matthew D. and Gelman, Andrew},
  title   = {The No-{U}-Turn Sampler: Adaptively Setting Path Lengths in Hamiltonian Monte Carlo},
  journal = {Journal of Machine Learning Research},
  year    = {2014},
  volume  = {15},
  number  = {47},
  pages   = {1593--1623}
}

@article{richardson1972rl,
  author  = {Richardson, William H.},
  title   = {Bayesian-Based Iterative Method of Image Restoration},
  journal = {Journal of the Optical Society of America},
  year    = {1972},
  volume  = {62},
  number  = {1},
  pages   = {55--59},
  doi     = {10.1364/JOSA.62.000055}
}

@article{lucy1974rl,
  author  = {Lucy, L. B.},
  title   = {An Iterative Technique for the Rectification of Observed Distributions},
  journal = {The Astronomical Journal},
  year    = {1974},
  volume  = {79},
  number  = {6},
  pages   = {745--754},
  doi     = {10.1086/111605}
}

@book{knoll2010rdm,
  author    = {Knoll, Glenn F.},
  title     = {Radiation Detection and Measurement},
  edition   = {4},
  year      = {2010},
  publisher = {Wiley},
  address   = {Hoboken, NJ}
}

@book{hansen2010rank,
  author    = {Hansen, Per Christian},
  title     = {Discrete Inverse Problems: Insight and Algorithms},
  series    = {Fundamentals of Algorithms},
  year      = {2010},
  publisher = {Society for Industrial and Applied Mathematics},
  address   = {Philadelphia, PA},
  doi       = {10.1137/1.9780898718836}
}

@book{tikhonov1977solutions,
  author    = {Tikhonov, Andrei N. and Arsenin, Vasiliy Y.},
  title     = {Solutions of Ill-Posed Problems},
  series    = {Scripta Series in Mathematics},
  year      = {1977},
  publisher = {V. H. Winston \& Sons},
  address   = {Washington, DC}
}

@book{kaipio2005statistical,
  author    = {Kaipio, Jari P. and Somersalo, Erkki},
  title     = {Statistical and Computational Inverse Problems},
  series    = {Applied Mathematical Sciences},
  volume    = {160},
  year      = {2005},
  publisher = {Springer},
  address   = {New York, NY},
  doi       = {10.1007/b138659}
}

@article{shepp1982em,
  author  = {Shepp, Lawrence A. and Vardi, Yehuda},
  title   = {Maximum Likelihood Reconstruction for Emission Tomography},
  journal = {IEEE Transactions on Medical Imaging},
  year    = {1982},
  volume  = {1},
  number  = {2},
  pages   = {113--122},
  doi     = {10.1109/TMI.1982.4307558}
}

@book{golub2013matrix,
  author    = {Golub, Gene H. and Van Loan, Charles F.},
  title     = {Matrix Computations},
  edition   = {4},
  series    = {Johns Hopkins Studies in the Mathematical Sciences},
  year      = {2013},
  publisher = {Johns Hopkins University Press},
  address   = {Baltimore, MD}
}

@book{engl1996regularization,
  author    = {Engl, Heinz Werner and Hanke, Martin and Neubauer, Andreas},
  title     = {Regularization of Inverse Problems},
  series    = {Mathematics and Its Applications},
  volume    = {375},
  year      = {1996},
  publisher = {Kluwer Academic Publishers},
  address   = {Dordrecht}
}

@article{vehtari2021rank,
  author  = {Vehtari, Aki and Gelman, Andrew and Simpson, Daniel and Carpenter, Bob and B{\"u}rkner, Paul-Christian},
  title   = {Rank-Normalization, Folding, and Localization: An Improved $\widehat{R}$ for Assessing Convergence of {MCMC}},
  journal = {Bayesian Analysis},
  year    = {2021},
  volume  = {16},
  number  = {2},
  pages   = {667--718},
  doi     = {10.1214/20-BA1221}
}

@article{arviz2019,
  author  = {Kumar, Ravin and Carroll, Colin and Hartikainen, Ari and Martin, Osvaldo},
  title   = {{ArviZ} a Unified Library for Exploratory Analysis of {Bayesian} Models in {Python}},
  journal = {Journal of Open Source Software},
  year    = {2019},
  volume  = {4},
  number  = {33},
  pages   = {1143},
  doi     = {10.21105/joss.01143}
}

@article{myllymaki2017global,
  author  = {Myllym{\"a}ki, Mari and Mrkvi{\v{c}}ka, Tom{\'a}{\v{s}} and Grabarnik, Pavel and Seijo, Henri and Hahn, Ute},
  title   = {Global Envelope Tests for Spatial Processes},
  journal = {Journal of the Royal Statistical Society: Series B (Statistical Methodology)},
  year    = {2017},
  volume  = {79},
  number  = {2},
  pages   = {381--404},
  doi     = {10.1111/rssb.12172}
}

@article{schiller2000extraction,
  author  = {Schiller, A. and Bergholt, L. and Guttormsen, M. and Melby, E. and Rekstad, J. and Siem, S.},
  title   = {Extraction of Level Density and $\gamma$ Strength Function from Primary $\gamma$ Spectra},
  journal = {Nuclear Instruments and Methods in Physics Research Section A},
  year    = {2000},
  volume  = {447},
  number  = {3},
  pages   = {498--511},
  doi     = {10.1016/S0168-9002(99)01187-0}
}

@article{toft2010sn,
  author  = {Toft, H. K. and Larsen, A. C. and Agvaanluvsan, U. and B{\"u}rger, A. and Guttormsen, M. and Mitchell, G. E. and Nyhus, H. T. and Schiller, A. and Siem, S. and Syed, N. U. H. and Voinov, A. V.},
  title   = {Level Densities and $\gamma$-Ray Strength Functions in {Sn} Isotopes},
  journal = {Physical Review C},
  year    = {2010},
  volume  = {81},
  number  = {6},
  pages   = {064311},
  doi     = {10.1103/PhysRevC.81.064311}
}

@article{zeiser2020oscar,
  author  = {Zeiser, F. and Tveten, G. M. and {Bello Garrote}, F. L. and Guttormsen, M. and Larsen, A. C. and Ingeberg, V. W. and G{\"o}rgen, A. and Siem, S.},
  title   = {The $\gamma$-Ray Energy Response of the {Oslo Scintillator Array OSCAR}},
  journal = {Nuclear Instruments and Methods in Physics Research Section A},
  year    = {2021},
  volume  = {985},
  pages   = {164678},
  doi     = {10.1016/j.nima.2020.164678}
}

@article{guttormsen2011168,
  author  = {Guttormsen, M. and B{\"u}rger, A. and Hansen, T. E. and Lietaer, N.},
  title   = {The {SiRi} Particle-Telescope System},
  journal = {Nuclear Instruments and Methods in Physics Research Section A},
  year    = {2011},
  volume  = {648},
  number  = {1},
  pages   = {168--173},
  doi     = {10.1016/j.nima.2011.05.055}
}

@article{myllymaki2024get,
  author  = {Myllym{\"a}ki, Mari and Mrkvi{\v{c}}ka, Tom{\'a}{\v{s}}},
  title   = {{GET}: Global Envelopes in {R}},
  journal = {Journal of Statistical Software},
  year    = {2024},
  volume  = {111},
  number  = {3},
  pages   = {1--40},
  doi     = {10.18637/jss.v111.i03}
}

@book{hansen1998rankdeficient,
  author    = {Hansen, Per Christian},
  title     = {Rank-Deficient and Discrete Ill-Posed Problems: Numerical Aspects of Linear Inversion},
  series    = {Mathematical Modeling and Computation},
  year      = {1998},
  publisher = {Society for Industrial and Applied Mathematics},
  address   = {Philadelphia, PA},
  doi       = {10.1137/1.9780898719697}
}

@article{skellam1946frequency,
  author  = {Skellam, J. G.},
  title   = {The Frequency Distribution of the Difference between Two {Poisson} Variates Belonging to Different Populations},
  journal = {Journal of the Royal Statistical Society},
  year    = {1946},
  volume  = {109},
  number  = {3},
  pages   = {296},
  doi     = {10.1111/j.2397-2335.1946.tb04670.x}
}

@misc{lima2025,
  author        = {Lima, E. and Braseth, L. L. and Mj{\o}s, A. H. and Hjorth-Jensen, M. and Kvellestad, A. and Larsen, A. C.},
  title         = {Regularized Unfolding of $\gamma$-Ray Spectra for Nuclear Physics Applications},
  year          = {2025},
  eprint        = {2511.16687},
  archivePrefix = {arXiv},
  primaryClass  = {physics.ins-det}
}

@article{pymc2023,
  author  = {Abril-Pla, Oriol and Andreani, Virgile and Carroll, Colin and Dong, Larry and Fonnesbeck, Christopher J. and Kochurov, Maxim and Kumar, Ravin and Lao, Junpeng and Luhmann, Christian C. and Martin, Osvaldo A. and Osthege, Michael and Vieira, Ricardo and Wiecki, Thomas and Zinkov, Robert},
  title   = {{PyMC}: A Modern and Comprehensive Probabilistic Programming Framework in {Python}},
  journal = {PeerJ Computer Science},
  year    = {2023},
  volume  = {9},
  pages   = {e1516},
  doi     = {10.7717/peerj-cs.1516}
}

@misc{phan2019numpyro,
  author        = {Phan, Du and Pradhan, Neeraj and Jankowiak, Martin},
  title         = {Composable Effects for Flexible and Accelerated Probabilistic Programming in {NumPyro}},
  year          = {2019},
  eprint        = {1912.11554},
  archivePrefix = {arXiv},
  primaryClass  = {stat.ML}
}

@misc{cabezas2024blackjax,
  author        = {Cabezas, Alberto and Corenflos, Adrien and Lao, Junpeng and Louf, R{\'e}mi and Carnec, Antoine and Chaudhari, Kaustubh and Cohn-Gordon, Reuben and Coullon, Jeremie and Deng, Wei and Duffield, Sam and Dur{\'a}n-Mart{\'i}n, Gerardo and Elantkowski, Marcin and Foreman-Mackey, Dan and Gregori, Michele and Iguaran, Carlos and Kumar, Ravin and Lysy, Martin and Murphy, Kevin and Orduz, Juan Camilo and Patel, Karm and Wang, Xi and Zinkov, Rob},
  title         = {{BlackJAX}: Composable {Bayesian} Inference in {JAX}},
  year          = {2024},
  eprint        = {2402.10797},
  archivePrefix = {arXiv},
  primaryClass  = {cs.MS}
}

@misc{seyboldt2026nutpie,
  author        = {Seyboldt, Adrian and Carlson, Eliot L. and Carpenter, Bob},
  title         = {Preconditioning Hamiltonian Monte Carlo by Minimizing Fisher Divergence},
  year          = {2026},
  eprint        = {2603.18845},
  archivePrefix = {arXiv},
  primaryClass  = {stat.CO}
}

@article{larsen201969,
  author  = {Larsen, A. C. and Spyrou, A. and Liddick, S. N. and Guttormsen, M.},
  title   = {Novel Techniques for Constraining Neutron-Capture Rates Relevant for $r$-Process Heavy-Element Nucleosynthesis},
  journal = {Progress in Particle and Nuclear Physics},
  year    = {2019},
  volume  = {107},
  pages   = {69--108},
  doi     = {10.1016/j.ppnp.2019.04.002}
}

@article{guttormsen1996371,
  author  = {Guttormsen, M. and Tveter, T. S. and Bergholt, L. and Ingebretsen, F. and Rekstad, J.},
  title   = {The Unfolding of Continuum $\gamma$-Ray Spectra},
  journal = {Nuclear Instruments and Methods in Physics Research Section A},
  year    = {1996},
  volume  = {374},
  number  = {3},
  pages   = {371--376},
  doi     = {10.1016/0168-9002(96)00197-0}
}

@article{kirsch201830,
  author  = {Kirsch, L. E. and Bernstein, L. A.},
  title   = {{RAINIER}: A Simulation Tool for Distributions of Excited Nuclear States and Cascade Fluctuations},
  journal = {Nuclear Instruments and Methods in Physics Research Section A},
  year    = {2018},
  volume  = {892},
  pages   = {30--40},
  doi     = {10.1016/j.nima.2018.02.096}
}

@misc{zeiser2020rainier,
  author       = {Kirsch, L. E. and Zeiser, Fabio},
  title        = {{fzeiser/RAINIER}: v1.4.1},
  year         = {2019},
  howpublished = {Zenodo},
  doi          = {10.5281/zenodo.2540342}
}

@misc{ompy_zenodo,
  author       = {Midtb{\o}, J{\o}rgen Eriksson and Zeiser, Fabio and Lima, Erlend and Ingeberg, Vetle W. and {Min RK}},
  title        = {{oslocyclotronlab/ompy}: {Ompy} v1.1.0},
  year         = {2020},
  howpublished = {Zenodo},
  doi          = {10.5281/zenodo.3898281}
}

@article{midtbo2021ompy,
  author  = {Midtb{\o}, J{\o}rgen E. and Zeiser, Fabio and Lima, Erlend and Larsen, Ann-Cecilie and Tveten, Gry M. and Guttormsen, Magne and {Bello Garrote}, Frank L. and Kvellestad, Anders and Renstr{\o}m, Therese},
  title   = {A New Software Implementation of the {Oslo} Method with Rigorous Statistical Uncertainty Propagation},
  journal = {Computer Physics Communications},
  year    = {2021},
  volume  = {262},
  pages   = {107795},
  doi     = {10.1016/j.cpc.2020.107795}
}

@misc{oclresponse2020,
  author       = {Zeiser, Fabio and Tveten, Gry M.},
  title        = {{oslocyclotronlab/OCL\_GEANT4}: {Geant4} Model of {OSCAR}},
  year         = {2020},
  howpublished = {Zenodo},
  note         = {Version v2.0.0.1},
  doi          = {10.5281/zenodo.4018494}
}

@article{dagostini1995unfolding,
  author  = {D'Agostini, G.},
  title   = {A Multidimensional Unfolding Method Based on {Bayes}' Theorem},
  journal = {Nuclear Instruments and Methods in Physics Research Section A},
  year    = {1995},
  volume  = {362},
  number  = {2--3},
  pages   = {487--498},
  doi     = {10.1016/0168-9002(95)00274-X}
}

@article{zech2013rl,
  author  = {Zech, G.},
  title   = {Iterative Unfolding with the {Richardson-Lucy} Algorithm},
  journal = {Nuclear Instruments and Methods in Physics Research Section A},
  year    = {2013},
  volume  = {716},
  pages   = {1--9},
  doi     = {10.1016/j.nima.2013.03.026}
}

@misc{choudalakis2012fbu,
  author        = {Choudalakis, Georgios},
  title         = {Fully {Bayesian} Unfolding},
  year          = {2012},
  eprint        = {1201.4612},
  archivePrefix = {arXiv},
  primaryClass  = {physics.data-an}
}

@article{kuusela2015unfolding,
  author  = {Kuusela, Mikael and Panaretos, Victor M.},
  title   = {Statistical Unfolding of Elementary Particle Spectra: Empirical {Bayes} Estimation and Bias-Corrected Uncertainty Quantification},
  journal = {The Annals of Applied Statistics},
  year    = {2015},
  volume  = {9},
  number  = {3},
  pages   = {1671--1705},
  doi     = {10.1214/15-AOAS857}
}

@article{diamond1980highspin,
  author  = {Diamond, R. M. and Stephens, F. S.},
  title   = {Nuclei at High Angular Momentum},
  journal = {Annual Review of Nuclear and Particle Science},
  year    = {1980},
  volume  = {30},
  number  = {1},
  pages   = {85--156},
  doi     = {10.1146/annurev.ns.30.120180.000505}
}

@article{radford1987stripping,
  author  = {Radford, D. C. and Ahmad, I. and Holzmann, R. and Janssens, R. V. F. and Khoo, T. L.},
  title   = {A Prescription for the Removal of Compton-Scattered Gamma Rays from Gamma-Ray Spectra},
  journal = {Nuclear Instruments and Methods in Physics Research Section A},
  year    = {1987},
  volume  = {258},
  number  = {1},
  pages   = {111--118},
  doi     = {10.1016/0168-9002(87)90086-6}
}

@article{love1989unfolding,
  author  = {Love, D. J. G. and Nelson, A. H.},
  title   = {Unfolding the Response Function of High-Quality Germanium Detectors},
  journal = {Nuclear Instruments and Methods in Physics Research Section A},
  year    = {1989},
  volume  = {274},
  number  = {3},
  pages   = {541--546},
  doi     = {10.1016/0168-9002(89)90188-5}
}

@article{koohifayegh1993neural,
  author  = {Koohi-Fayegh, R. and Green, S. and Crout, N. M. J. and Taylor, G. C. and Scott, M. C.},
  title   = {Neural Network Unfolding of Photon and Neutron Spectra Using an {NE-213} Scintillation Detector},
  journal = {Nuclear Instruments and Methods in Physics Research Section A},
  year    = {1993},
  volume  = {329},
  number  = {1--2},
  pages   = {269--276},
  doi     = {10.1016/0168-9002(93)90946-F}
}

@misc{mjos2026archive,
  author       = {Mj{\o}s, Andreas H. and others},
  title        = {Empirical-Bayes Unfolding of Gamma-Ray Spectra:
                  Code and Results},
  year         = {2026},
  publisher    = {Zenodo},
  version      = {1.0.0},
  doi          = {10.5281/zenodo.20797045}
}

\end{document}